\documentclass[final]{ustcstep}

\usepackage[dvips]{graphicx}
\usepackage[square]{natbib}

\def\gsim{\;\lower4pt\hbox{${\buildrel\displaystyle >\over\sim}$}\;}
\def\lsim{\;\lower4pt\hbox{${\buildrel\displaystyle <\over\sim}$}\;}
\def\grls{\;\lower4pt\hbox{${\buildrel\displaystyle >\over <}$}\;}

\newcommand\addr[2]{{\footnotesize \it $^{#1}$#2}\\}

\begin{document}

\title{Statistical Study of Coronal Mass Ejection Source Locations: II. Role of Active Regions in CME Production}

\author{Caixia Chen,$^1$ Yuming Wang,$^{1,*}$ Chenglong Shen,$^1$ Pinzhong Ye,$^1$ Jie Zhang,$^2$ and S. Wang $^1$\\[1pt]
\addr{1}{CAS Key Lab of Geospace Environment, Department of Geophysics \& Planetary
Sciences, University of Science \& Technology}
\addr{ }{ of China, Hefei, Anhui 230026, China}
\addr{2}{School of Physics, Astronomy and
Computational  Sciences, George Mason University 4400 University
Dr., MSN 6A2, Fairfax,}
\addr{ }{ VA 22030, USA}
\addr{*}{To whom correspondence should be addressed. E-mail: ymwang@ustc.edu.cn}}

\maketitle
\tableofcontents

\begin{abstract}
This is the second paper of the statistical study of coronal mass ejection (CME) source locations,
in which the relationship between CMEs and active regions (ARs) is statistically studied on the basis
of the information of CME source locations and the ARs automatically extracted from magnetic synoptic 
charts of Michelson Doppler Imager (MDI) during 1997 -- 1998. Totally, 224 CMEs with a known location
and 108 MDI ARs are included in our sample. It is found that about 63\% of the CMEs are related
with ARs, at least about 53\% of the ARs produced one or more CMEs, and particularly about 14\% of ARs
are CME-rich (3 or more CMEs were generated) during one transit across the visible disk. Several issues are then tried to clarify: whether or 
not the CMEs originating from ARs are distinct from others, whether or not the
CME kinematics depend on AR properties, and whether or not the CME productivity depends on AR properties.
The statistical results suggest that (1) there is no evident difference between AR-related and 
non-AR-related CMEs in terms of CME speed, acceleration and width, (2) the size, strength and complexity
of ARs do little with the kinematic properties of CMEs, but have significant effects on
the CME productivity, and (3) the sunspots in all the most productive ARs at least belong to $\beta\gamma$
type, whereas 90\% of those in CME-less ARs are $\alpha$ or $\beta$ type only. A detailed analysis on 
CME-rich ARs further reveals that (1) the distribution of the waiting time of same-AR CMEs,
consists of two parts with a separation at about 15 hours, which implies that the CMEs with a waiting time shorter 
than 15 hours are probably truly physical related, and (2) an AR tends to
produce such related same-AR CMEs at a pace of 8 hours, but cannot produce two or more fast CMEs 
($>800$ km s$^{-1}$) within a time interval of 15 hours. This interesting phenomenon is particularly discussed.
\end{abstract}

\section{Introduction}
Coronal mass ejections (CMEs) are one of the most violent explosive
phenomena in the solar atmosphere, and active regions (ARs) are
thought to be the most efficient producer of CMEs because free
energy tends to  accumulate there. However, different ARs may have
different capability of generating CMEs, and CMEs may not be
necessary to take place in ARs. These two facts leave the
relationship between CMEs and ARs still an unresolved issue.

Previous studies have shed  light on the AR's capability of
producing (strong) CMEs.  Through examining 117 ARs,
\citet{Canfield_etal_1999} found that ARs are  more likely to be
eruptive if they are either sigmoidal or large.
\citet{Guo_etal_2007} investigated 55 flare-CME productive ARs and
found that fast CMEs tended to initiate in ARs with large magnetic
flux or long lengths of main polarity inversion lines (PILs). Through investigating 57 fastest
CMEs with speed larger than 1500 km s$^{-1}$ from 1996 June to 2007
January as well as 1143 ARs recognized from magnetic synoptic charts
obtained by Michelson Doppler Imager (MDI) on board Solar and
Heliospheric Observatory (SOHO), \citet{Wang_Zhang_2008} found that
there was a general trend that a larger, stronger, and more complex
AR was more likely to produce a faster CME. A systematical study was
also performed by \citet{Falconer_etal_2002, Falconer_etal_2006,
Falconer_etal_2008, Falconer_etal_2009} in their series papers. They
found that the CME productivity of a bipolar AR depended on the
global nonpotentiality of the AR's magnetic field. Furthermore,
\citet{Yeates_etal_2010} identified and investigated 98 front-side
CMEs during 1999 May 13 -- September 26, compared their source
regions with the simulation results of coronal magnetic field evolution,
and found that the strong gradient of the radial component of magnetic
field at photosphere, that usually appears in ARs, may be a good 
indicator of CME-productive regions.

Similar dependence on AR free energy  can be found in many studies
of the flare productivity of ARs \citep[e.g.,][]{Sammis_etal_2000,
Leka_Barnes_2003b, Leka_Barnes_2007, Maeshiro_etal_2005,
Jing_etal_2006, Ternullo_etal_2006, Schrijver_2007,
Georgoulis_Rust_2007, Su_etal_2007}. Although flares are also a
violent explosive phenomenon in the solar atmosphere, they are
different from CMEs.  Flares  can be classified as either confined
ones or  eruptive ones according to whether or not they are
associated  with CMEs \citep[e.g.,][]{Svestka_Cliver_1992,
Wang_Zhang_2007, Schrijver_2009}. Thus the statistical results
obtained for flares and CMEs are similar but not the same. An
example for the difference between flares and CMEs can be seen from
the flare and CME productivities of an AR-complex reported by
\citet{Akiyama_etal_2007}, in which two adjacent flare-productive
ARs have much different levels of CME association. Moreover, they
found that for the CME-rich AR, the average waiting time of flares
is much longer than that for the CME-poor AR. We know that
sufficient free energy is a necessary condition for an AR to be
eruptive \citep[e.g.,][]{Priest_Forbes_2002, Regnier_Priest_2007}.
Since both flares and CMEs  consume the free energy, flares and CMEs
sometimes may work as two competing  processes. From this
perspective, to understand  AR's ability of producing  CMEs is
different from that producing flares, and thus becomes a more
complicated issue.

On the other hand, the association of CMEs with ARs has also been
widely studied. Through examining 32 CMEs whose source regions were
located on the solar disk and well observed in EIT 195 \AA\ from
1996 January through 1998 May, \citet{Subramanian_Dere_2001} found
that about 84\% CMEs were associated with ARs.
\citet{Zhou_etal_2003} studied 197 front-side halo CMEs (angular
width $> 130^\circ$) from 1997 to 2001 and found that there were
about 79\% front-side halo CMEs originating from ARs. It has been
suggested  for a long time that there might  be two distinct types
of CMEs \citep[e.g.,][]{MacQueen_Fisher_1983, StCyr_etal_1999,
Sheeley_etal_1999, Delannee_etal_2000, Andrews_Howard_2001,
Moon_etal_2002}. One type of CMEs is associated with flares and
usually originates from ARs; they have a constant or decreasing speed
in the outer corona, implying an impulsive acceleration process 
in the inner corona. The other type of CMEs is often associated
with quiescent filament-eruptions; their speeds increase with a
nearly constant acceleration, implying a gradual acceleration
process. However, several more recent statistical studies reached
an opposite conclusion that there is no two distinct types of CMEs 
\citep[e.g.,][]{Yurchyshyn_etal_2005, Vrsnak_etal_2005, Chen_etal_2006b}.
Counter cases can be often observed. For example, \citet{Feynman_Ruzmaikin_2004} 
presented a quiescent filament-associated CME, which reached
an extremely fast speed in the corona. Similar cases can be found in
the paper by \citet{Wang_Zhang_2008}, e.g., the CMEs occurring on
1998 April 20 and 2002 May 22. Thus, the issue whether or not there are two
distinct types of CMEs and the role of ARs in this issue are 
worth to be clarified.

Apparently, further studies are needed to fully understand the role
of ARs in producing CMEs.   What kind of ARs can or cannot produce
CMEs? What kind of ARs can frequently produce CMEs?  What causes
different kinematic properties of CMEs? Any inputs from
observations, in particular, results from statistical studies,  can
be used to constrain theoretical models. In our previous study
\citep[][hereafter referred as Paper I]{Wang_etal_2011}, we have
manually identified the source locations of all  CMEs from 1997 to
1998, and a total of 288 CMEs have been located their source regions
on the visible solar disk (refer to
\url{http://space.ustc.edu.cn/dreams/cme_sources/}). In our another
paper by \citet{Wang_Zhang_2008}, we developed an automatic method
to detect and quantitatively characterize ARs from photospheric
magnetogram images. Thus, the two works provide us the
observational base   for investigating the relationship between ARs
and CMEs. The paper is organized as follows. In Sec.\ref{sec_data},
we introduce the data of CMEs and ARs which will be used in this
study. Then we present the statistical  results of the dependence of
CME apparent properties on ARs in Sec.\ref{sec_dep}.  The  CME
productivity of  ARs is presented  in Sec.\ref{sec_productivity}.
In Sec.\ref{sec_CME-rich}, we further study
 those ARs frequently producing CMEs. Finally, summary and
conclusions are given in Sec.\ref{sec_conclusions} and Sec.\ref{sec_discussion}.

\section{Data and Method}\label{sec_data}
\begin{figure*}[tbh]
  \centering
  \includegraphics[width=\hsize]{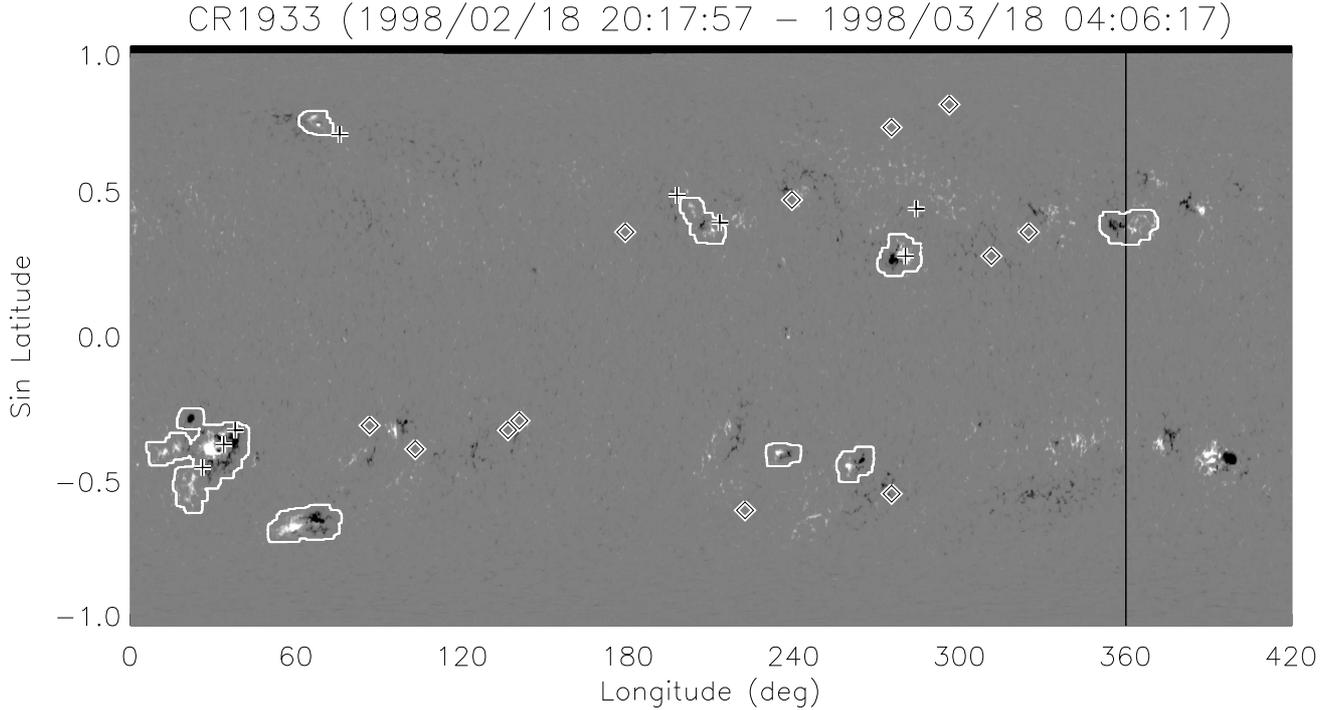}
  \caption{ MDI magnetic synoptic chart of Carrington rotation 1933. A small portion on the right-most side is from previous Carrington rotation.
  Extracted MDI ARs are marked by the enclosing white lines. Plus symbols represent the locations of
   AR-related CMEs,  while diamonds indicate the locations of  non-AR-related CMEs.}\label{fg_carr}
\end{figure*}

ARs usually appear as bright patches on the Sun in the EUV
wavelengths, and have strong magnetic field. A frequently referred
catalog of ARs is compiled by NOAA SWPC\footnote{Space Weather
Prediction Center,
\url{http://www.swpc.noaa.gov/ftpmenu/forecasts/SRS.html}}, in which
several parameters of ARs and the corresponding  sunspot groups are
given, such as the location, area, classifications, sunspot number,
etc. However, the NOAA AR catalog  lacks of some key quantitative
information of ARs such as magnetic field strength, flux, etc. For
this sake, we developed an automatic method in 2008 to extract ARs
based on the synoptic charts of photospheric magnetic field from
SOHO/MDI; they are called MDI ARs. Through
this method, ARs can be recognized and parameterized with  a uniform
set of criteria,  free of personal biases in the identification
process. A detailed description of the method and the comparison
of  MDI ARs with NOAA ARs can be found in \citet{Wang_Zhang_2008}
and a follow-up paper by \citet{Zhang_etal_2010}.

In this paper, we will use the MDI ARs rather than the traditional
NOAA ARs to study the role of ARs in producing CMEs.
Figure~\ref{fg_carr} shows the MDI ARs from Carrington rotation
1933, as an example. The plus and diamond symbols  marked on the
map indicate the locations of AR-related and non-AR-related CMEs,
respectively; the Carrington longitude and latitude of these CMEs
correspond to the heliographic coordinates of the CME source
location  at the time observed in EIT.

To determine if a CME is related to an AR and which AR is related
to, we first identify the source locations of the CME. As mentioned 
before, all the LASCO CMEs during 1997 -- 1998 had been checked with
their source locations, and 288 CMEs were identified as front-side 
CMEs, namely location identified (LI) CMEs. One can refer to Paper I 
Section 2 for the detailed process of 
the identification. Briefly, we manually checked SOHO/EIT movies, and 
looked for any surface signatures of CMEs, such as flares, dimmings, 
waves, post-eruption loops, etc. If there was one or several eruption 
signatures reasonably close to the time and direction of a CME viewed 
in SOHO/LASCO, the CME is considered as a LI CME, and the center of 
the surface eruption feature is then chosen as its location.

Then we calculate the spherical surface distances ($D_{AR}$, in units 
of degree) between the CME and the boundaries of nearby ARs. If 
there is at least one AR within a threshold distance
$D_{AR}^{th}$, the CME is AR-related (as marked by the pluses in
Fig.~\ref{fg_carr}) and the related AR is the one having the
shortest distance; otherwise, the CME is non-AR-related (the
diamonds in Fig.~\ref{fg_carr}). Considering the error in
determination of CME locations and the projection effect for those
CMEs close to solar limb, we set $D_{AR}^{th}=5^\circ$ for CMEs with
$DSC<0.85R_S$ and $D_{AR}^{th}=10^\circ$ for CMEs with $DSC \geq
0.85R_S$. Here the quantity $DSC$ is the projected distance on the
plane-of-sky between the CME location and the solar disk center (see
Paper I for details). Meanwhile, for each AR, we classify it as
either a CME-less or CME-producing AR, depending on whether a CME is
associated with this AR or not.  Further, we define an AR as a
CME-rich AR, if it produced  three or more CMEs.

Compared to a snapshot MDI  magnetogram image, a synoptic chart does
not   show the exact state of the photospheric magnetic field  during a
CME.  However, it has the advantage of reduced
projection effect, in particular, for those CMEs far away from the
solar disk center.  For these CMEs, it is almost impossible to
 obtain the  correct information of photospheric magnetic field surrounding the
CME source location,  due to the presence of significant projection
effect. Further,  snapshot magnetograms cannot  provide us the
magnetic information behind the solar limb.  On the other hand, as
shown in Paper I there were 56\% of  CMEs with known source location
occurring for $DSC\geq0.85R_S$. Thus, it is necessary  to use
MDI synoptic charts for the study of this paper.

\begin{table}[bt]
\begin{center}
\caption{Numbers of different types of CMEs and MDI ARs}
\label{tb_ars}
\tabcolsep 23pt
\begin{tabular}{rcc}
\hline
\multicolumn{3}{c}{CMEs}\\
AR-related & 141  & 63\% \\
Non-AR-related & 83  & 37\% \\
Total & 224 & \\
\hline
\multicolumn{3}{c}{MDI ARs}\\
CME-less & 51 & 47\% \\
CME-producing$^1$ & 57 &53\%\\
CME-rich$^2$ & 15 &14\% \\
Total & 108 & \\
\hline
\end{tabular}
\end{center}
\footnotesize
$^1$ A CME-producing AR means the AR produced at least one CME during 
its passage across the visible disk.\\
$^2$ A CME-rich AR means the AR produced 3 or more CMEs. Thus CME-rich
ARs are a subset of CME-producing ARs.
\end{table}

Due to the presence of data gaps of SOHO observations, some MDI 
magnetic synoptic charts are incomplete. The CMEs corresponding to these
incomplete synoptic charts are simply excluded in the analysis. 
Also some LI CMEs with a low confidence level ($CL=3$) are removed.
Finally, there are in total 224 LI CMEs with  MDI synoptic
charts available and a total of 108 MDI ARs during the period of
study from 1997 -- 1998. It is straightforward to obtain that
 about 63\% of LI CMEs are related with ARs, while  the rest 37\% of LI CMEs are  not related with any  AR. Meanwhile,
 about 47\% of ARs do  not produce a single CME during the period crossing through the
visible solar disk.  About 53\% of ARs  produce at least one CME.
Particularly, about 14\% of ARs
 produce at least 3 CMEs, thus are CME-rich. The numbers of different types of CMEs and MDI ARs are
summarized in Table~\ref{tb_ars}. The fractions of different types
of MDI ARs are not accurate, because we are unable to learn the
activity of an AR before it  rotates to the front-side of the disk
and after it  rotates to the back-side of disk. Nevertheless, one
could assume that the activity level of a particular AR, is similar
in the front-side as in the back-side. It is probably true as one
will see in Sec.\ref{sec_rich-less} that the CME productivity of ARs is related
with the AR complexity, but not the AR phase.

\begin{figure}[tb]
  \centering
  \includegraphics[width=\hsize]{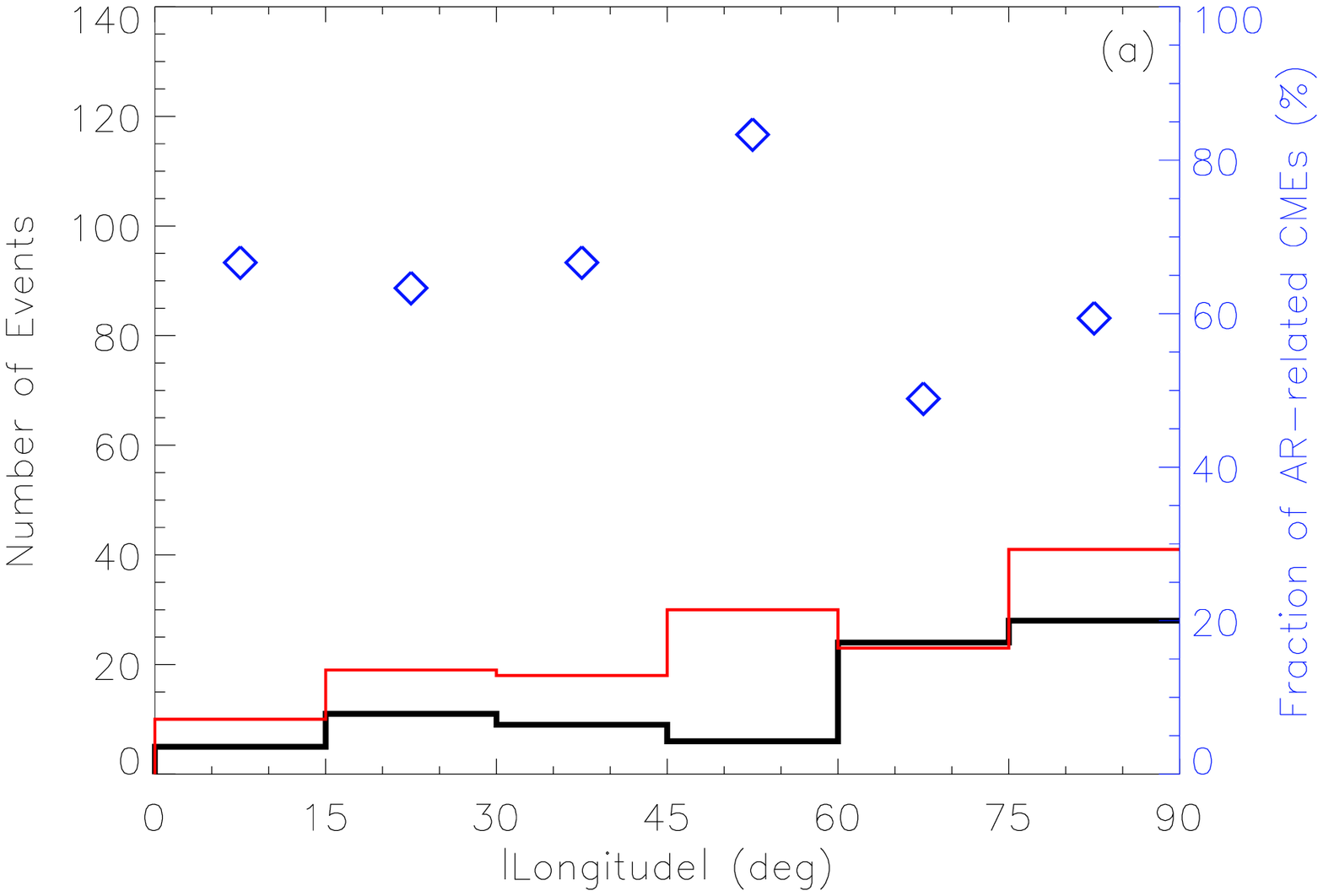}
\includegraphics[width=\hsize]{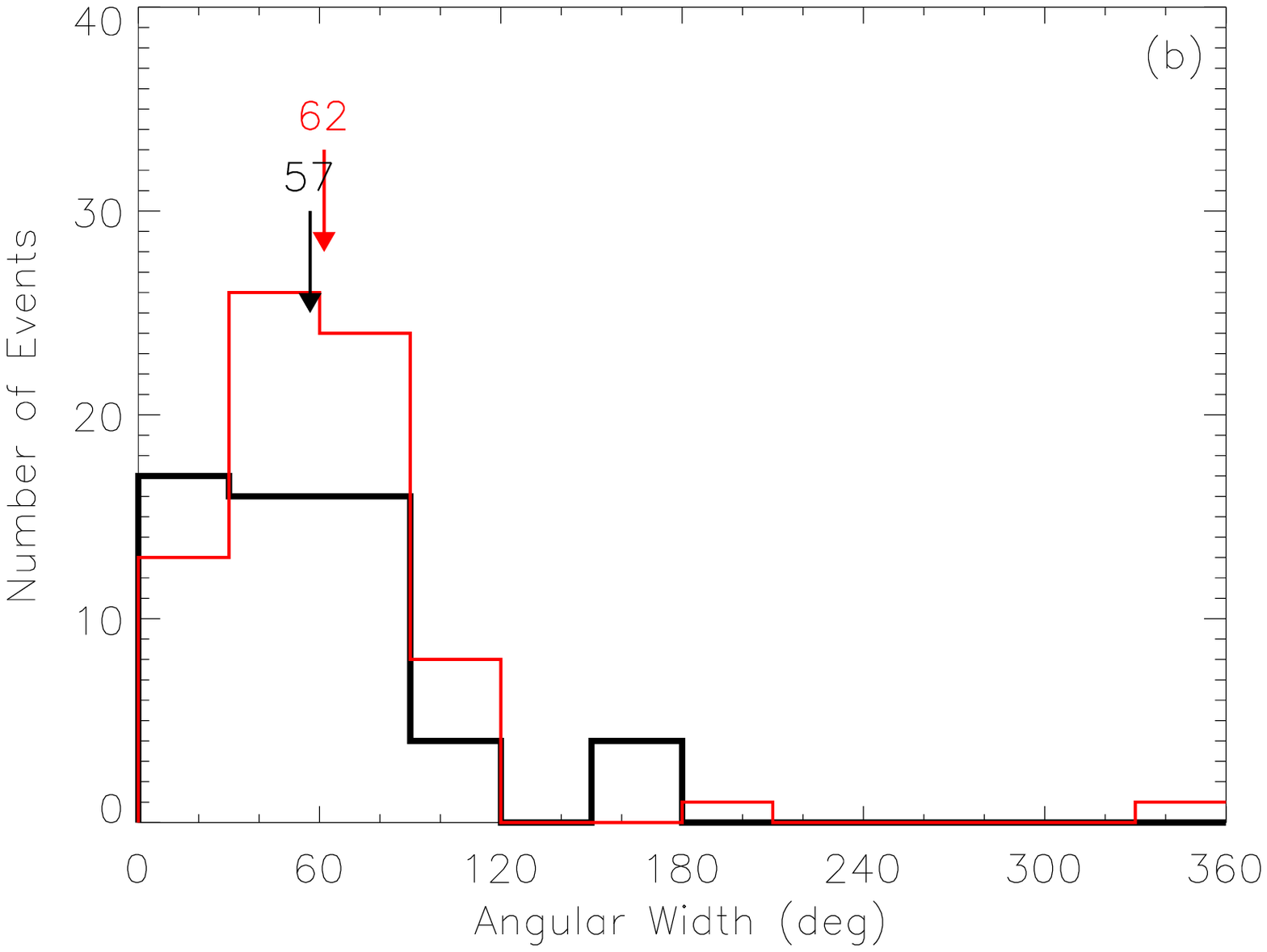}
  \caption{Distributions of AR-related (red) and non-AR-related (black) CMEs along absolute
longitude (panel a) and apparent angular width (panel b). In panel a, diamonds indicate the 
fraction of AR-related CMEs, which is scaled by the right $y$-axis. In panel b, on-disk CMEs 
are not included, and the digital numbers with arrows mark the average values.}\label{fg_long}
\end{figure}

\section{Dependence of CME Apparent Properties on ARs}\label{sec_dep}

\begin{figure}[tb]
  \centering
  \includegraphics[width=\hsize]{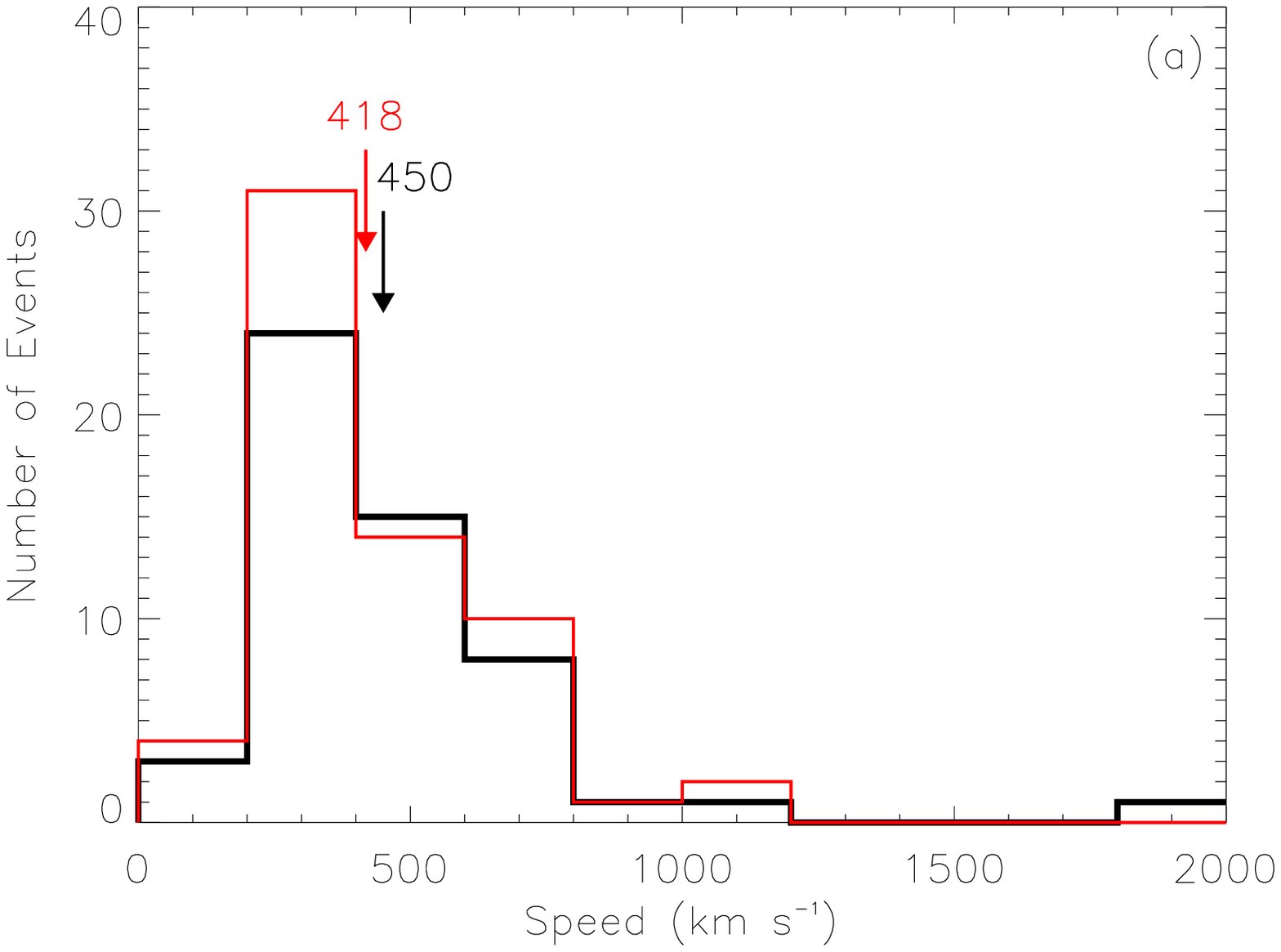}
  \includegraphics[width=\hsize]{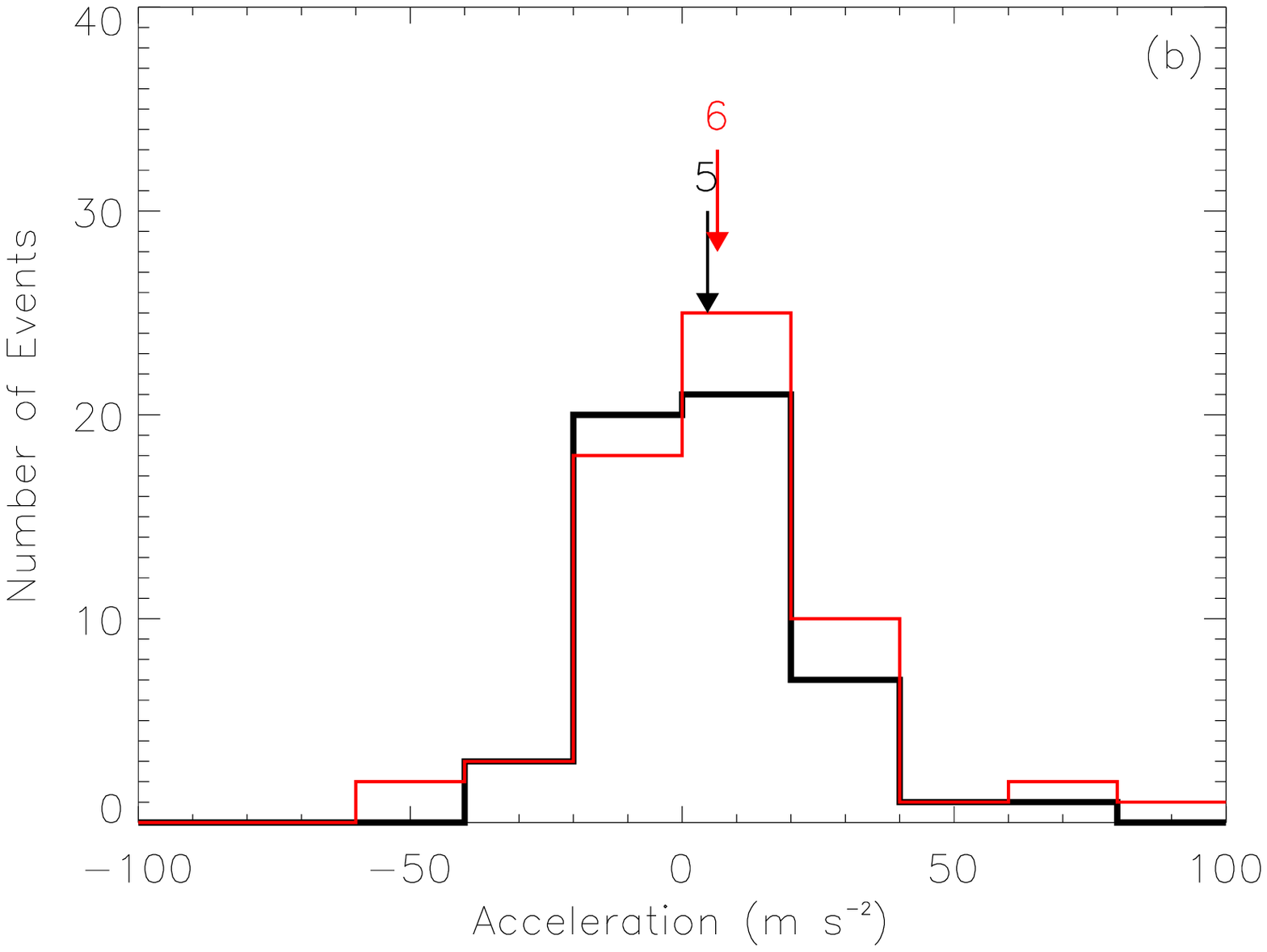}
  \caption{The histograms of the apparent speed (panel a) and
  acceleration (panel b) for AR-related (red) and non-AR-related (black) limb CMEs, respectively.
  The digital numbers with arrows mark the average values.
}\label{fg_arcme}
\end{figure}

First of all, we make a comparison study of  AR-related and
non-AR-related CMEs. The association rate of CMEs with ARs is about
63\% in this study. The variation of association rate along the
absolute value of the 
heliographic longitude is shown in Figure \ref{fg_long}a, in which
one can find that there is no significant difference between the limb
and on-disk fraction of AR-related CMEs. Thus we can conclude that the associations
of limb CMEs with ARs are reliable even though the projection effect
is maximized in determining the source location of limb CMEs. 
Moreover, the fraction of AR-related CMEs is decreasing only slightly for longitude $>60^\circ$.
Since $DSC$ and longitude are closely related for low latitudes (where ARs are located),
this justifies the simple criteria used for AR association (see the 4th paragraph of Sec.\ref{sec_data}). 
In particular, the sudden increase of $D_{AR}^{th}$ from $5^\circ$ to $10^\circ$ has not the effect to 
increase the CME association to ARs for $DSC \geq 0.85 R_S$.
But there is a significant increase of both numbers of AR-related and 
non-AR-related CMEs with longitude. This is due to the presence of occulting
effect, Thomson scattering effect and projection effect (see Paper I for details). 

The value of the association rate, 63\%, obtained in this study is smaller than 84\% and 79\% obtained
respectively by \citet{Subramanian_Dere_2001} and
\citet{Zhou_etal_2003}. This difference seems to be  caused by the
bias in the  selection of events. In their studies, only well
observed CMEs and/or
 halo CMEs were investigated, while in this paper, all CMEs are included,  
no matter whether a CME  is halo or narrow, and bright
or faint. This difference  suggests that there is a significant
fraction of CMEs may originate from quiet Sun regions, and these
CMEs tend to be weak and/or narrow. Figure \ref{fg_long}b 
presents the distribution of the apparent angular width for AR-related 
and non-AR-related CMEs with $DSC\geq 0.85R_S$, in which the projection 
effect is minimized. A weak difference
could be found between the two sets of CMEs that the
non-AR-related CMEs are slightly narrower than AR-related CMEs.

As mentioned in the Introduction, there perhaps exist two types of 
CMEs in terms of their kinematic behavior. One type of CMEs is 
impulsive and often associated with flares, and the other type of 
CMEs is gradual and often associated with prominences. The former 
type of CMEs usually has a faster speed and smaller acceleration 
in the outer corona than the latter \citep[e.g.,][]{Sheeley_etal_1999}.
Here, we compare the AR-related and non-AR-related CMEs, in order to 
check whether or not there are two different types of CMEs caused by 
difference types of source regions.

\begin{figure}[tb]
  \centering
  \includegraphics[width=\hsize]{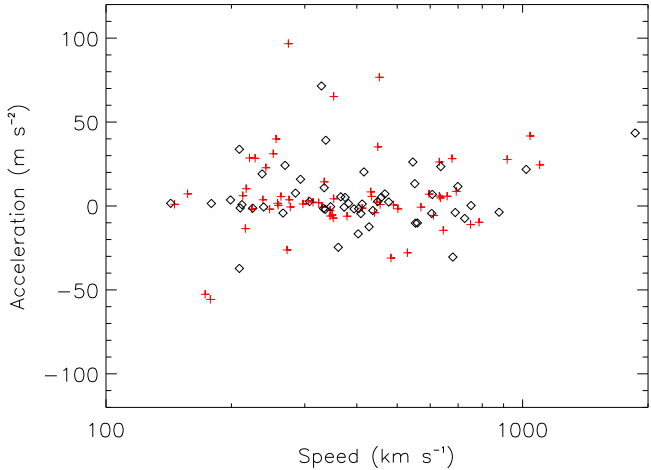}
  \caption{Acceleration versus the speed for the AR-related (pluses)
and non-AR-related (diamonds) limb CMEs.}\label{fg_speedacc}
\end{figure}

Figure~\ref{fg_arcme} shows the distributions of apparent
speed and acceleration of the AR-related and
non-AR-related CMEs. To minimize the bias of  the projection
effect, only limb CMEs (i.e., $DSC\geq0.85R_S$ and width
$<360^\circ$) with effectively measured speed and acceleration are considered here. This selection results in  62
AR-related CMEs and 53 non-AR-related CMEs.  As shown in the figure,
the distributions of the two sets of CMEs are quite similar. Both
AR-related and non-AR-related CMEs can reach a very fast speed
and/or a large acceleration/deceleration. Further, we show
the scattering plot between CME speeds and accelerations for the 
two sets of CMEs in Figure \ref{fg_speedacc}.
There is no evident difference between the two sets of CMEs. These
results are consistent with the studies by \citet{Yurchyshyn_etal_2005,
Vrsnak_etal_2005, Chen_etal_2006b}, who applied different classifications 
and also found no evidence supporting the existence of two distinct types of CMEs.

Second, we investigate  if the AR properties  may have an influence
on the CMEs kinematic properties.  We again consider only limb CMEs,
to reduce the projection effect;  full halo CMEs and those CME
without  speed measured are removed from our sample. There are 71
AR-related limb CMEs originating from 42 ARs. Since some ARs
produced more than one CME, the CME number is more than the AR
number. For those multiple-CME-producing ARs, we use the fastest CME
as the representative of the AR in the following analysis,  because
the fastest CME may reflect the capability of an AR producing a
strong eruption.

For each MDI AR, our automatic AR-detection method can extract at
least 12 parameters, including  that of areas, magnetic fluxes,
magnetic field strength, AR shape and PILs. We choose the following parameters for  further analysis:
total area ($A_t$), total magnetic flux ($F_t$), total length of
PILs ($L_{pil}$) and number of PILs ($N_{pil}$)\footnote{In our algorithm of recognizing 
AR and extracting parameters, some pixels in an AR with weak magnetic field are removed 
due to the preset threshold \citep[refer to][]{Wang_Zhang_2008}. The present threshold perhaps may
also remove some pixels around PILs so that positive and negative polarities may be no longer 
apparently adjacent and PILs can not be extracted. 
Actually, this treatment may keep main PILs and ignore minor PILs. Thus, 
ARs without PILs do exist in our sample, but they are not unipolar regions.}. 
These parameters
had proved to have influence on  AR's capability of producing
extremely fast CMEs \citep[see][]{Wang_Zhang_2008}.

\begin{table}[tb]
\begin{center}
\caption{Results of the linear regression analysis} \label{tb_deps}
\tabcolsep 5.5pt
\begin{tabular}{ccccccc}
\hline
&$c_0$ &$c_1$ &$c_2$ &$c_3$ &$c_4$ &$cc$ \\
\hline
Speed &453.81 &-44.66 &-20.78 &68.31 &2.23 &0.22 \\
Width & 58.23 &-39.09 &35.62 &15.97 &-11.49 &0.45 \\
\hline
\end{tabular}
\end{center}
\footnotesize
$^*$ Column $c_{0-4}$ are the coefficients in Eq.\ref{eq_1}. The last column gives the 
correlation coefficients between the observed values and the fitting results from 
the linear regression analysis. The second and third row is for the CME apparent
speed and width, respectively.
\end{table}

\begin{figure*}[p]
  \centering
  \includegraphics[width=0.4\hsize]{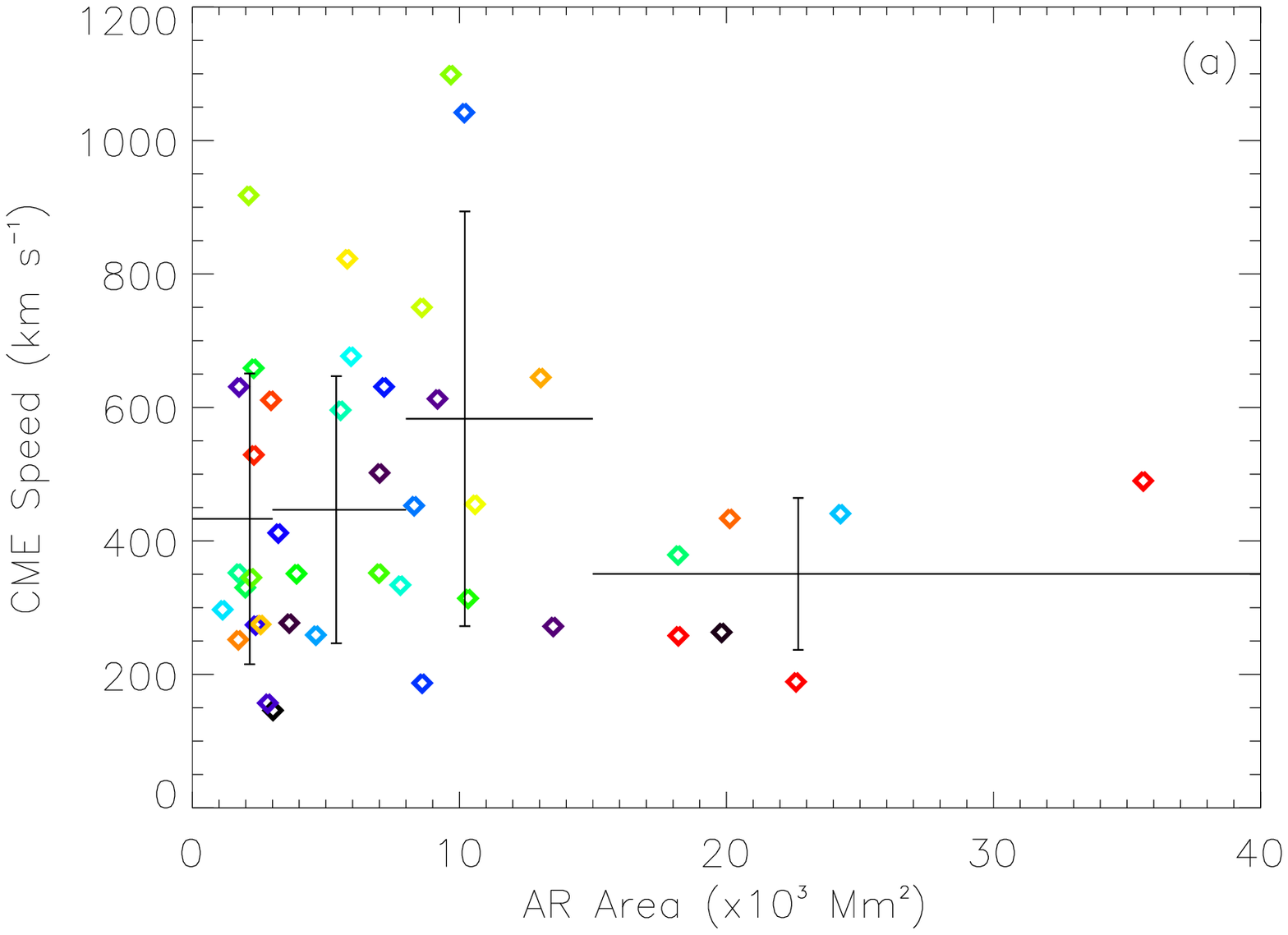}
  \includegraphics[width=0.4\hsize]{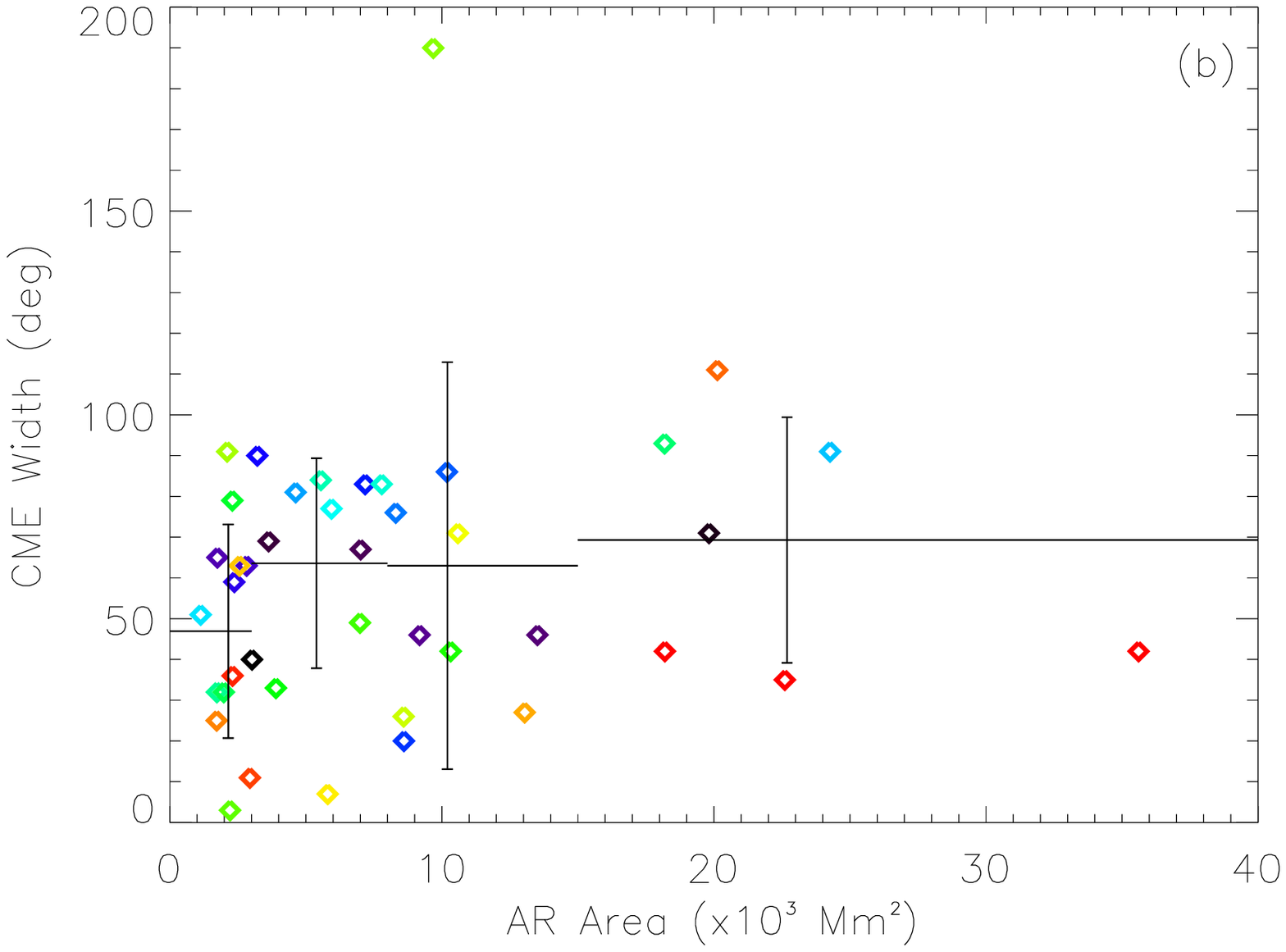}\\
  \includegraphics[width=0.4\hsize]{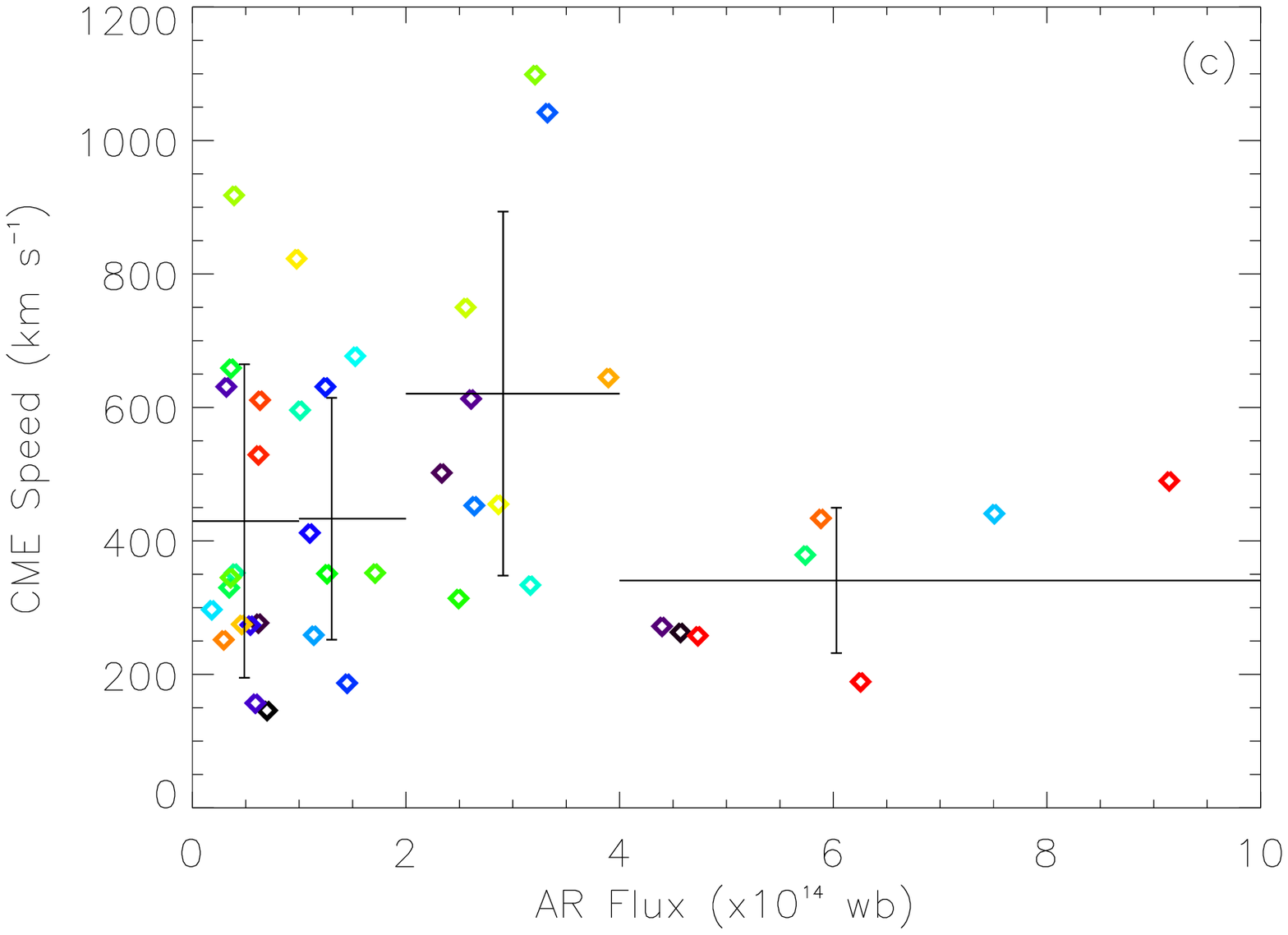}
  \includegraphics[width=0.4\hsize]{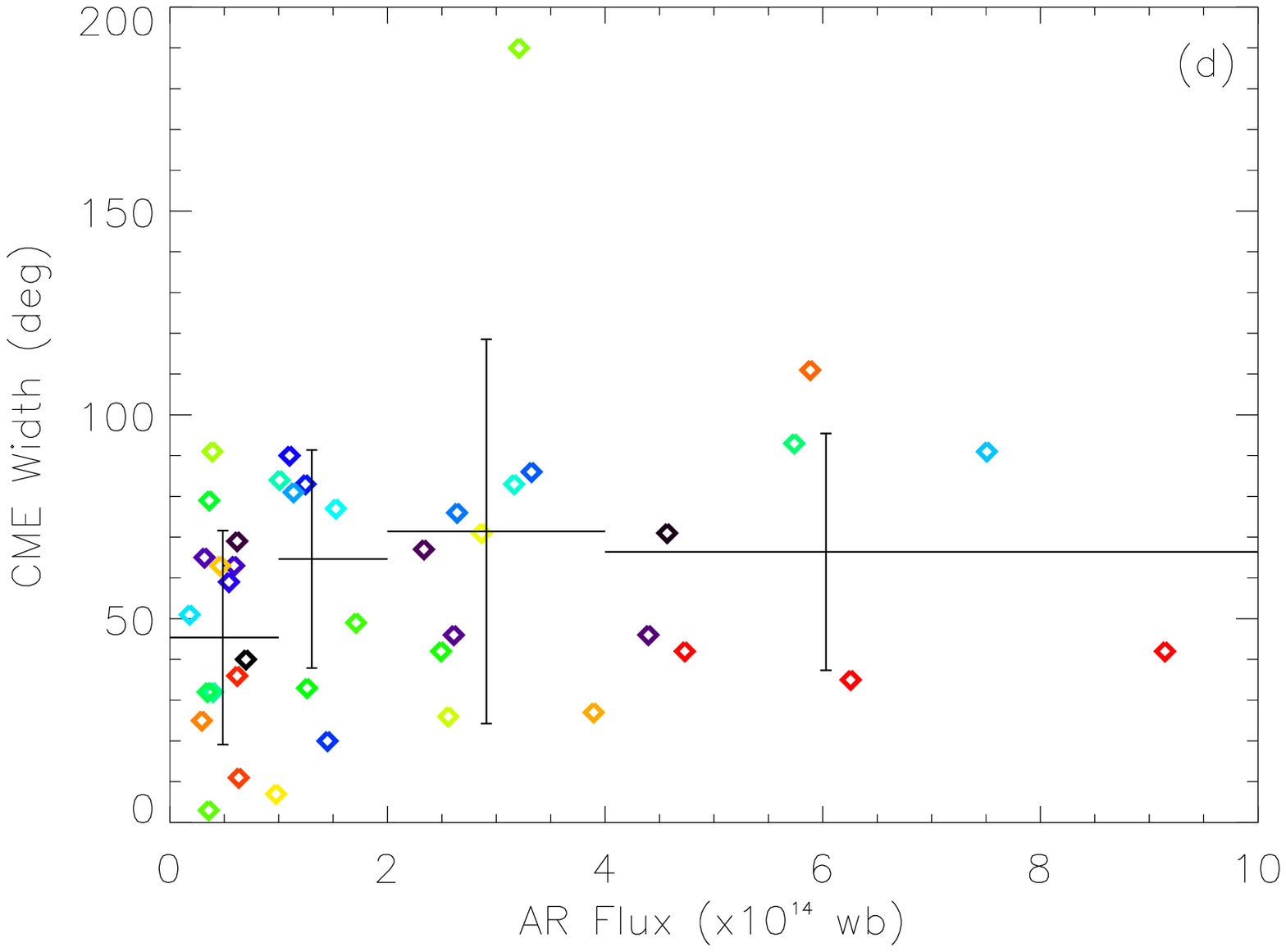}\\
  \includegraphics[width=0.4\hsize]{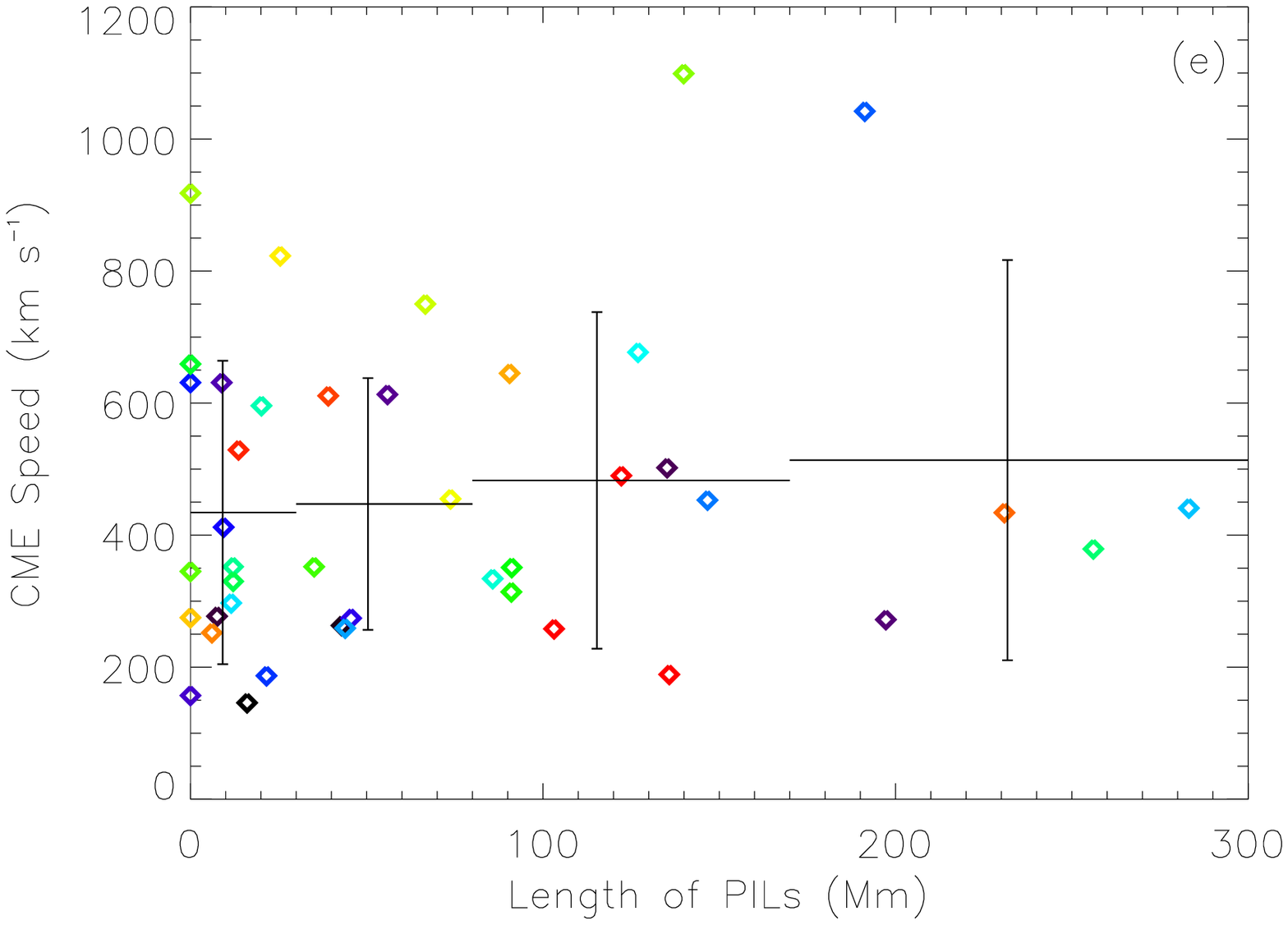}
  \includegraphics[width=0.4\hsize]{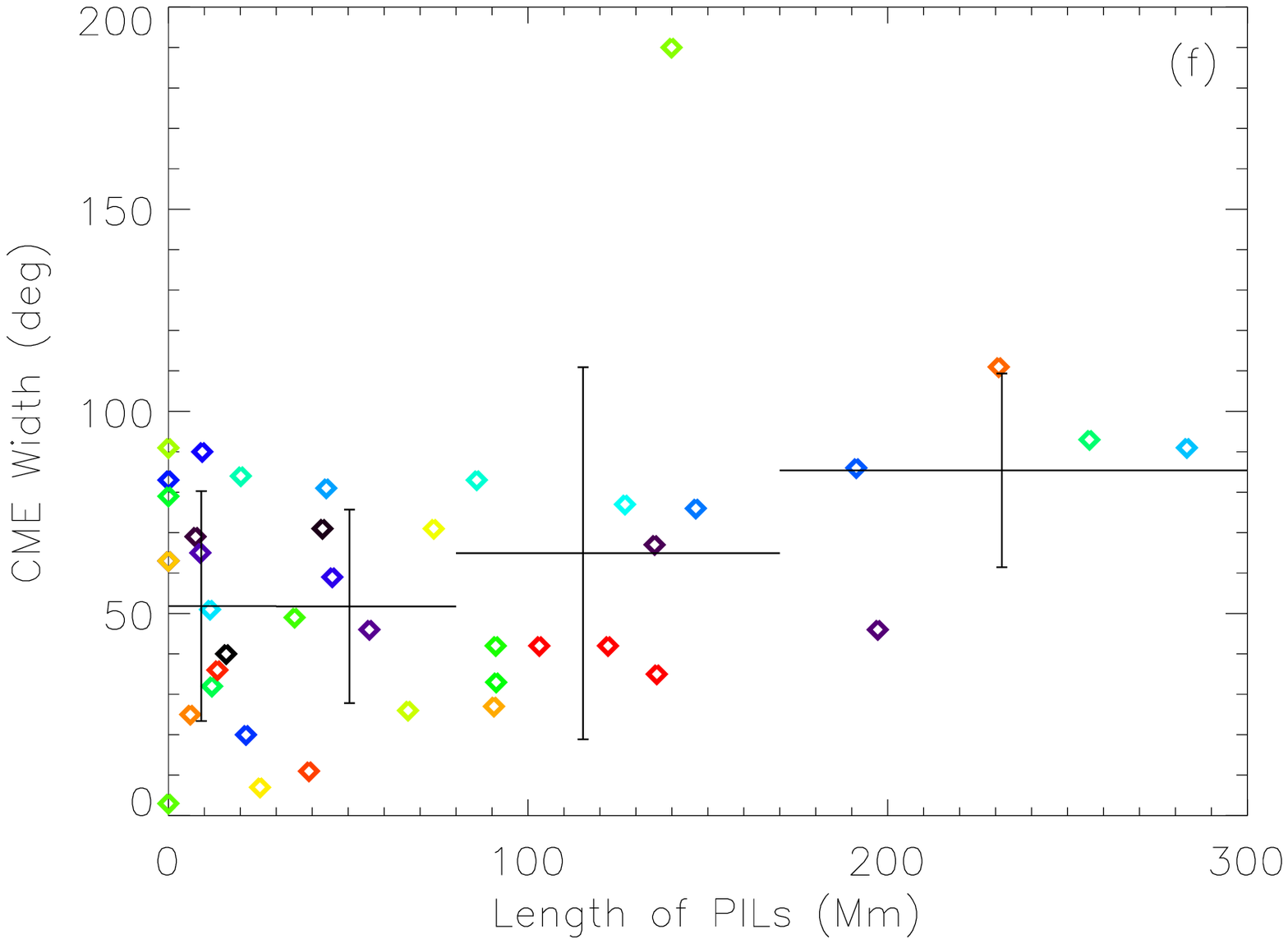}\\
  \includegraphics[width=0.4\hsize]{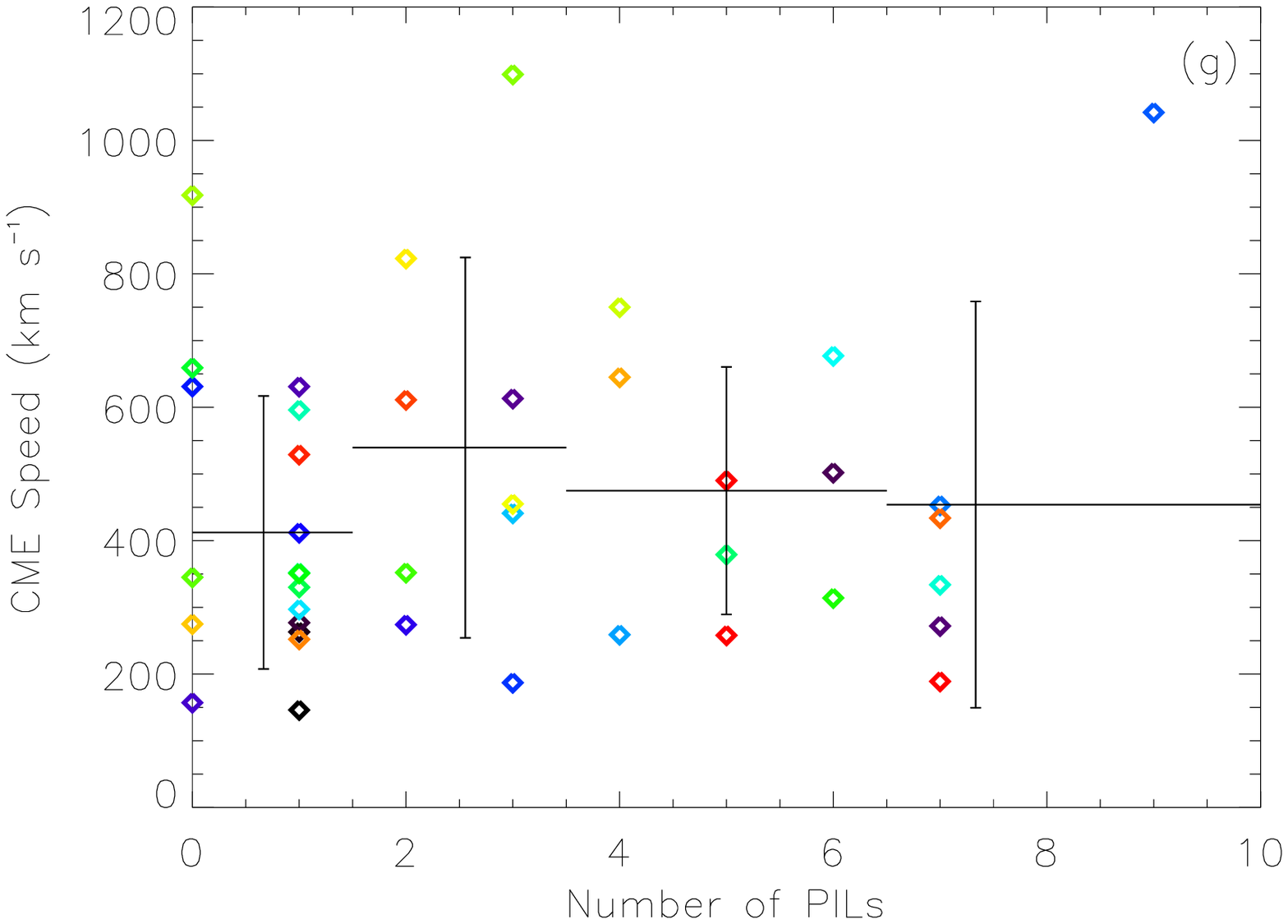}
  \includegraphics[width=0.4\hsize]{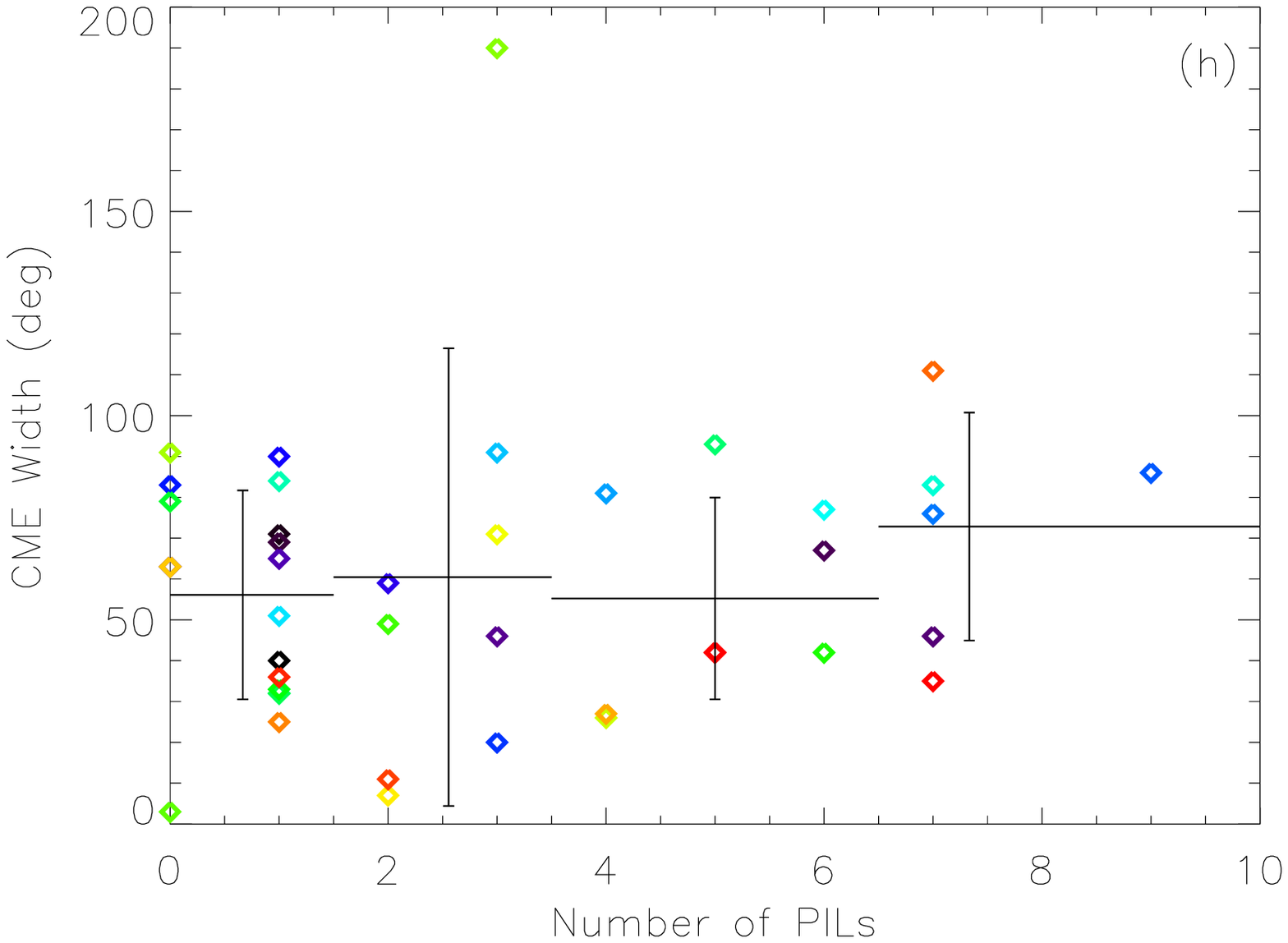}\\
  \caption{Scattering plots showing the possible correlation between the CME
parameters and  source AR parameters. The panels on the left  are
for CME  apparent speed and the panels on  the right  are for CME apparent angular width.
From the top to bottom, the panels are for AR area, magnetic flux,
length of PILs and  number of PILs, respectively. The data points 
are color coded just for one's convenience to compare the relative positions of
each data point in all the 8 sub-figures. The plus symbols
in each plot mark the average values of the data points within the
bin size indicated  by the horizontal extension of the symbol;  the
vertical extension of the plus symbols  indicate the standard
deviation of the data points. }\label{fg_dep}
\end{figure*}

Figure~\ref{fg_dep} presents the dependence of  CME speeds and
angular widths on  four AR parameters: $A_t$, $F_t$, $L_{pil}$ and
$N_{pil}$. In each panel, the plus symbols mark the average value
and the standard deviation of the data points within
 the range indicated by the horizontal bars. Apparently, no evident
correlation  can be found for these parameters.  We further look
into the possibility  that CME speed and width may be correlated
with the combination of the AR parameters. Thus, we apply  linear
regression analysis on the data.  The following function is fitted:
\begin{eqnarray}
y&=&c_0+c_1\frac{A_t}{<A_t>}+c_2\frac{F_t}{<F_t>}+c_3\frac{L_{pil}}{<L_{pil}>}\nonumber\\
&&+c_4\frac{N_{pil}}{<N_{pil}>} \label{eq_1}
\end{eqnarray}
where $y$ is the CME speed or angular width, $<x>$ means the average
value of  quantity $x$, and $c_{0-4}$ are the coefficients to be
fitted. 

\begin{figure}[tb]
  \centering
  \includegraphics[width=\hsize]{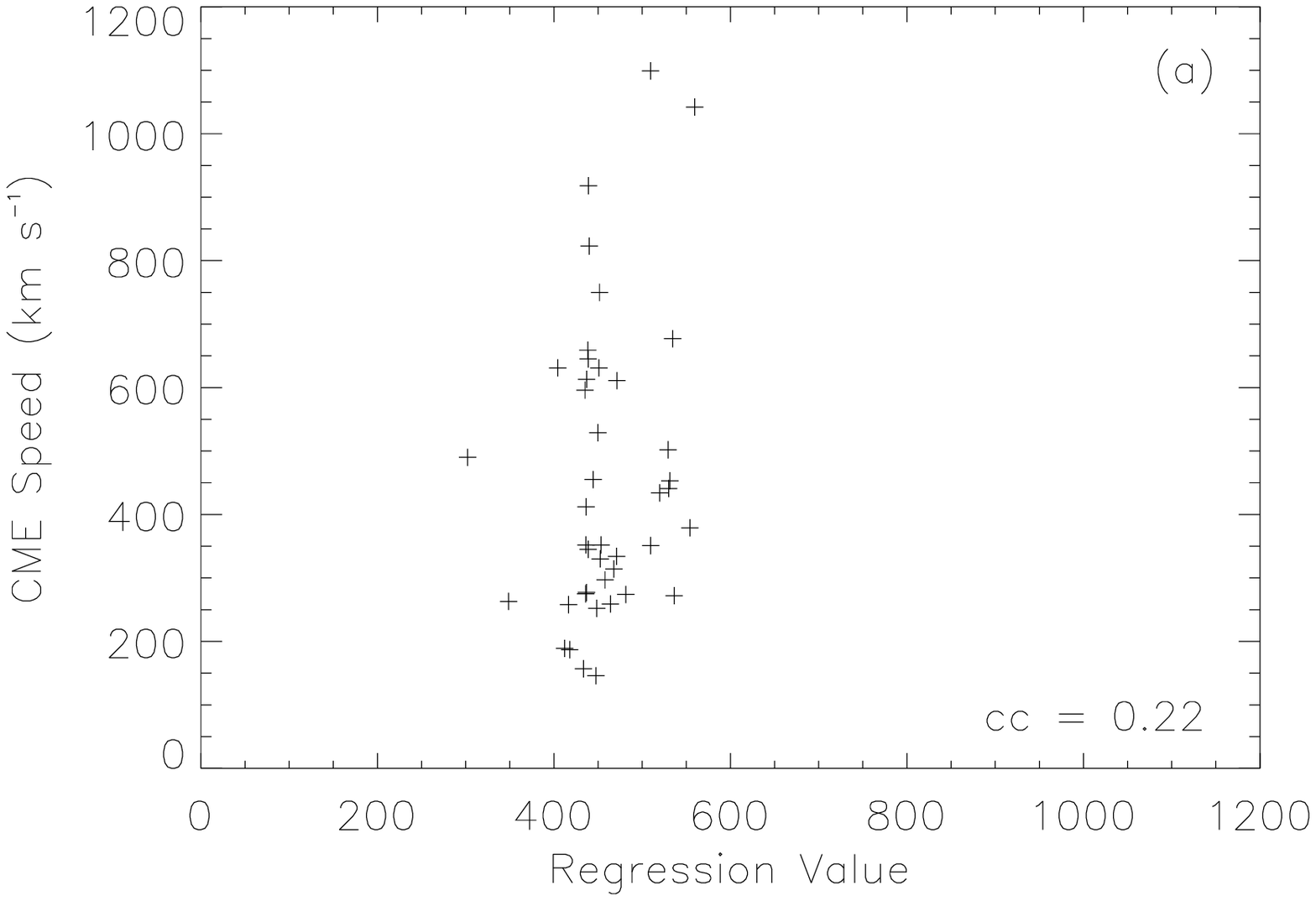}
  \includegraphics[width=\hsize]{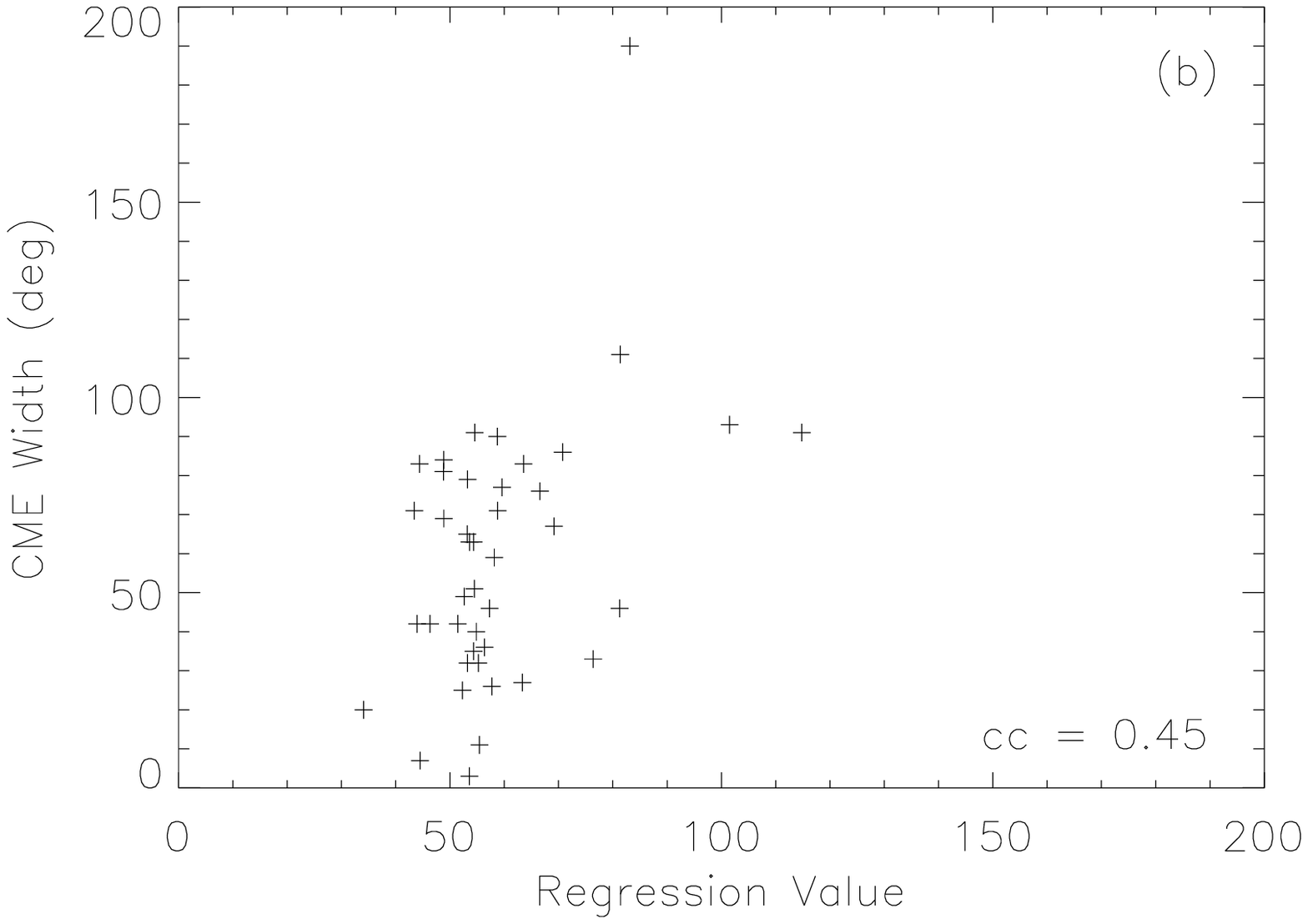}
  \caption{Correlation plots of the measured values versus linear regression values for CME apparent speeds (panel a) and widths (panel b).}\label{fg_deps}
\end{figure}

Figure~\ref{fg_deps} shows the fitting results, and the
obtained coefficients, $c_{0-4}$ and correlation coefficient, $cc$,
are listed in Table~\ref{tb_deps}. For CME speed, the value of $cc$
is only 0.22,  suggesting that there is almost no correlation
between  CME speed and the AR parameters we chose. In our previous
study \citep{Wang_Zhang_2008}, we  reached a conclusion that an AR
with larger area, stronger magnetic field and more complex
morphology has a higher possibility of  producing  extremely fast
CMEs (speed $>1500$ km s$^{-1}$). Our statistical result in this
paper  indicates that the same  conclusion cannot be extended to
CMEs with slower speed. For CME width, a weak correlation
($cc=0.45$) can be seen in Figure~\ref{fg_deps}(b).  It means that
the size, strength and complexity of ARs  may have an impact on the
size of produced CMEs. Moreover, from Table~\ref{tb_deps}, we find
that the coefficients, $c_1$ and $c_2$, are most significant,
suggesting that AR area and total magnetic flux are  more important
factors in determining the CME size.

\section{CME Productivity of ARs}\label{sec_productivity}

\begin{figure}[tb]
  \centering
  \includegraphics[width=\hsize]{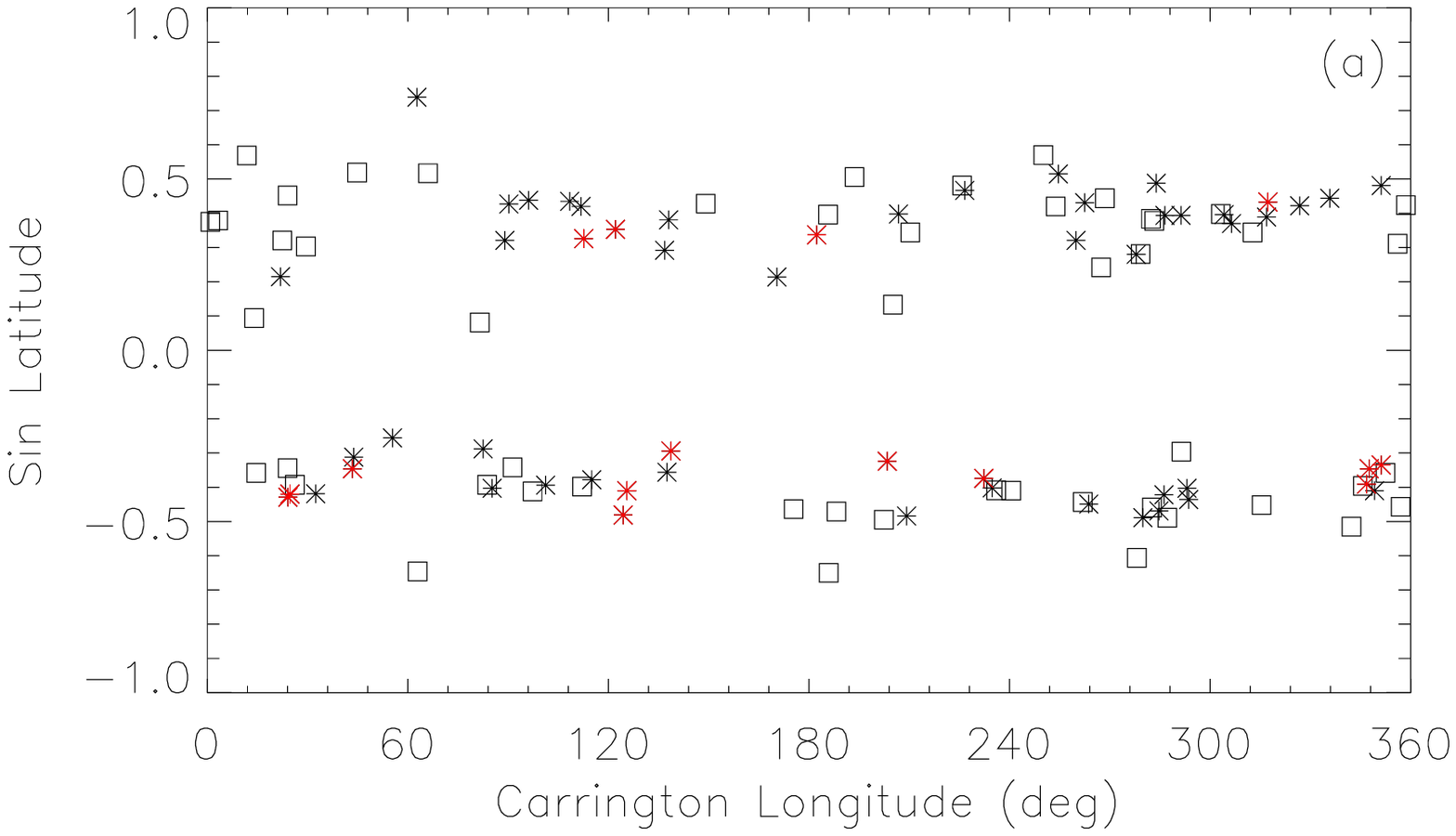}
  \includegraphics[width=\hsize]{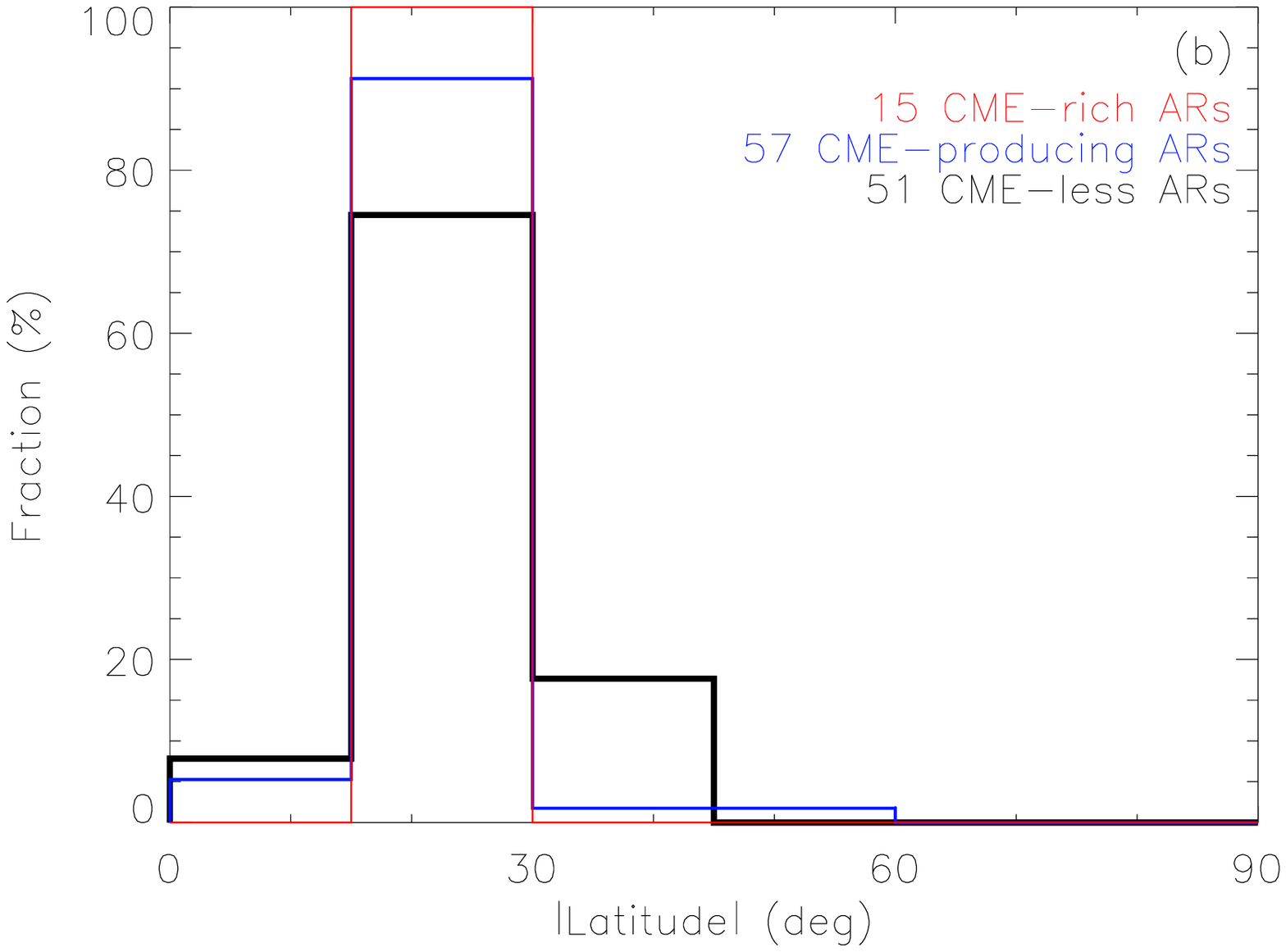}
  \caption{Panel (a): The distributions of the central locations (Carrington longitude and sine latitude)
  of CME-less (square symbols), CME-producing (asterisks) and CME-rich (red asterisks) ARs. Panel (b):
The histograms of the latitudes of the CME-less (black),
CME-producing (blue) and CME-rich (red) ARs.}\label{fg_dis}
\end{figure}

CMEs may originate from either ARs or quiet Sun regions. Reversely,
ARs may frequently produce CMEs or may not produce even a single
one. Why do different ARs have different CME productivity? This
issue is investigated by comparing  CME-less, CME-producing and
CME-rich ARs. Figure~\ref{fg_dis} shows the distribution of the
heliographic location (measured from the geometric center) of the
108 MDI ARs studied. The CME-less, CME-producing and CME-rich ARs
are indicated in different symbols or colors (see the figure
caption). All  MDI ARs appeared within latitude of $\pm60^\circ$,
and 83\% of them are located in two belts between latitude of
$\pm(15^\circ-30^\circ)$. Although the overall distributions of the
different types of ARs are quite similar, there is still certain
weak difference between them, which can be seen in
Figure~\ref{fg_dis}(b). For the CME-less ARs, there are about 25\% of
them occurring  outside of the two AR belts.  In contrast, all the
CME-rich ARs locate in the two belts. Consequently,  only 18\% of
ARs outside the two belts can produce CMEs.

\begin{figure*}[tbh]
  \centering
  \includegraphics[width=0.495\hsize]{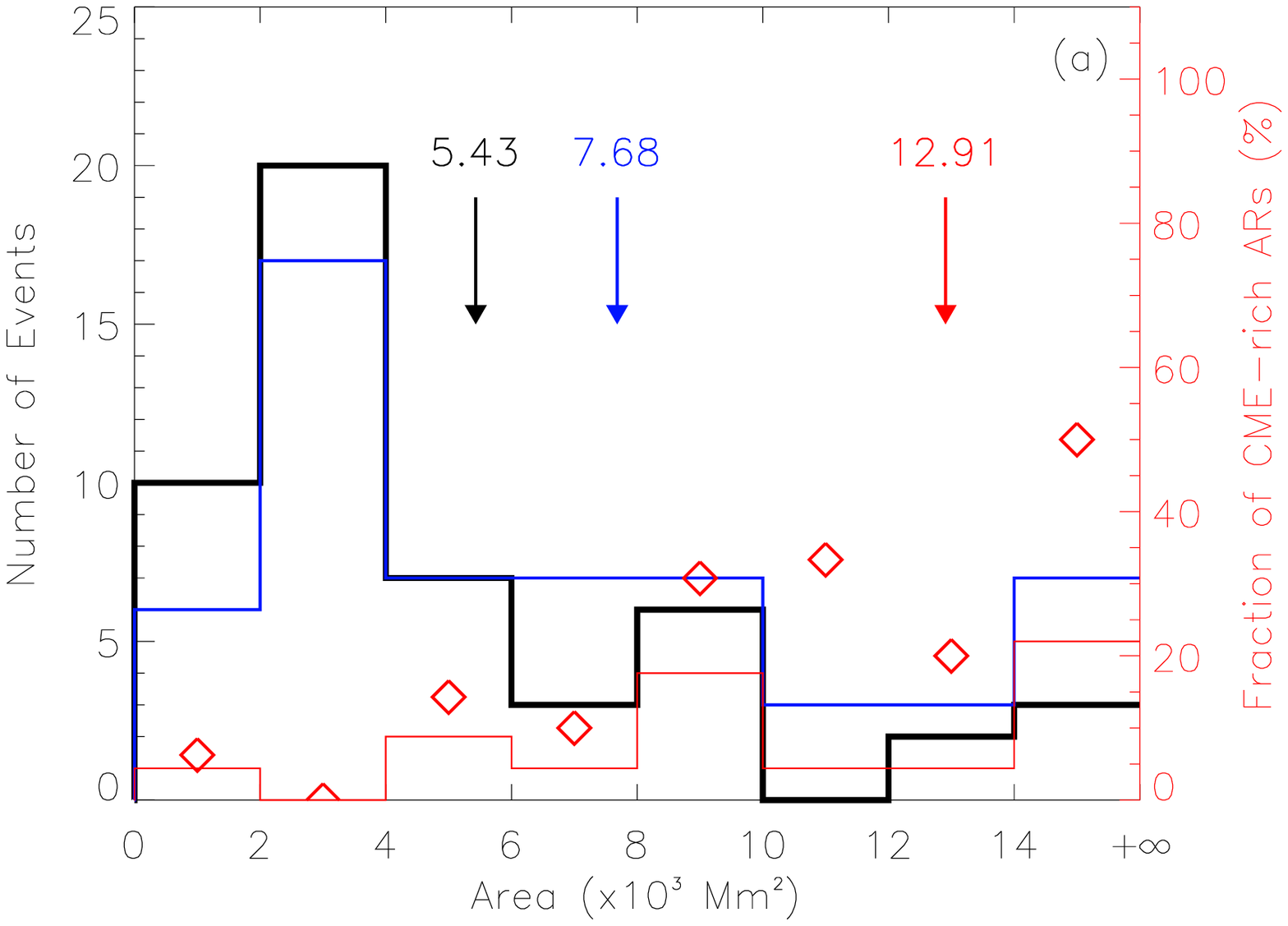}
  \includegraphics[width=0.495\hsize]{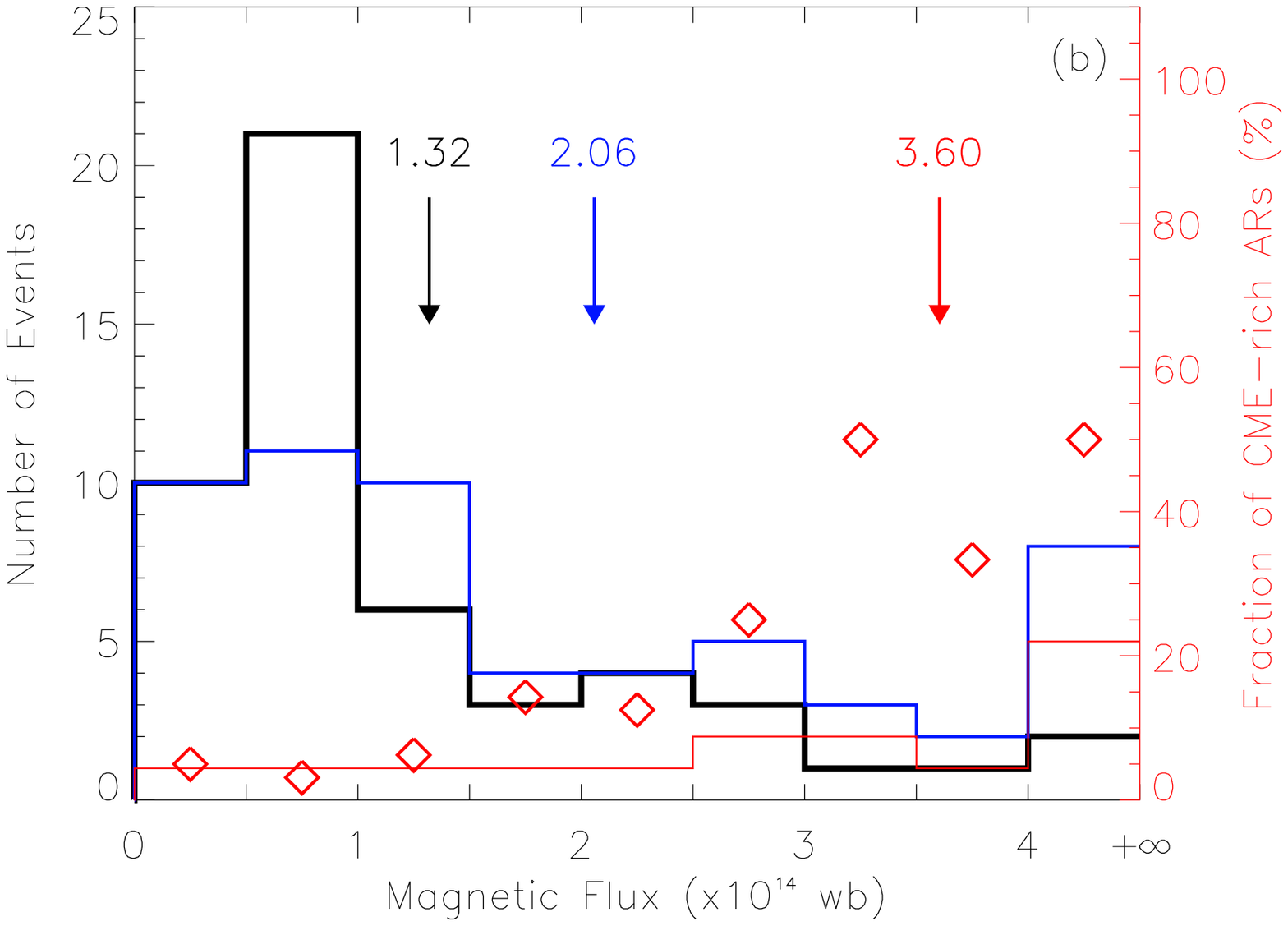} \\
  \includegraphics[width=0.495\hsize]{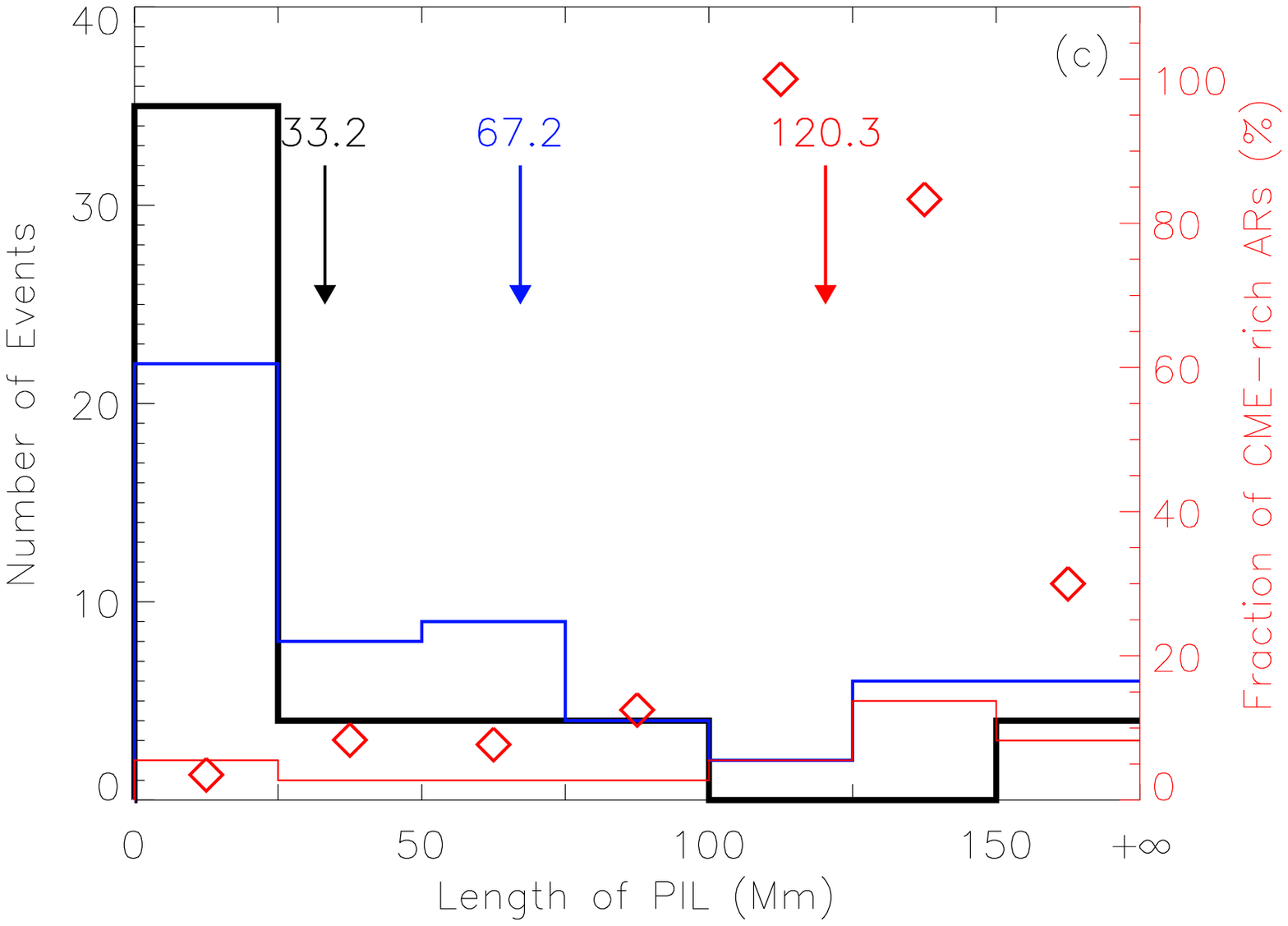}
  \includegraphics[width=0.495\hsize]{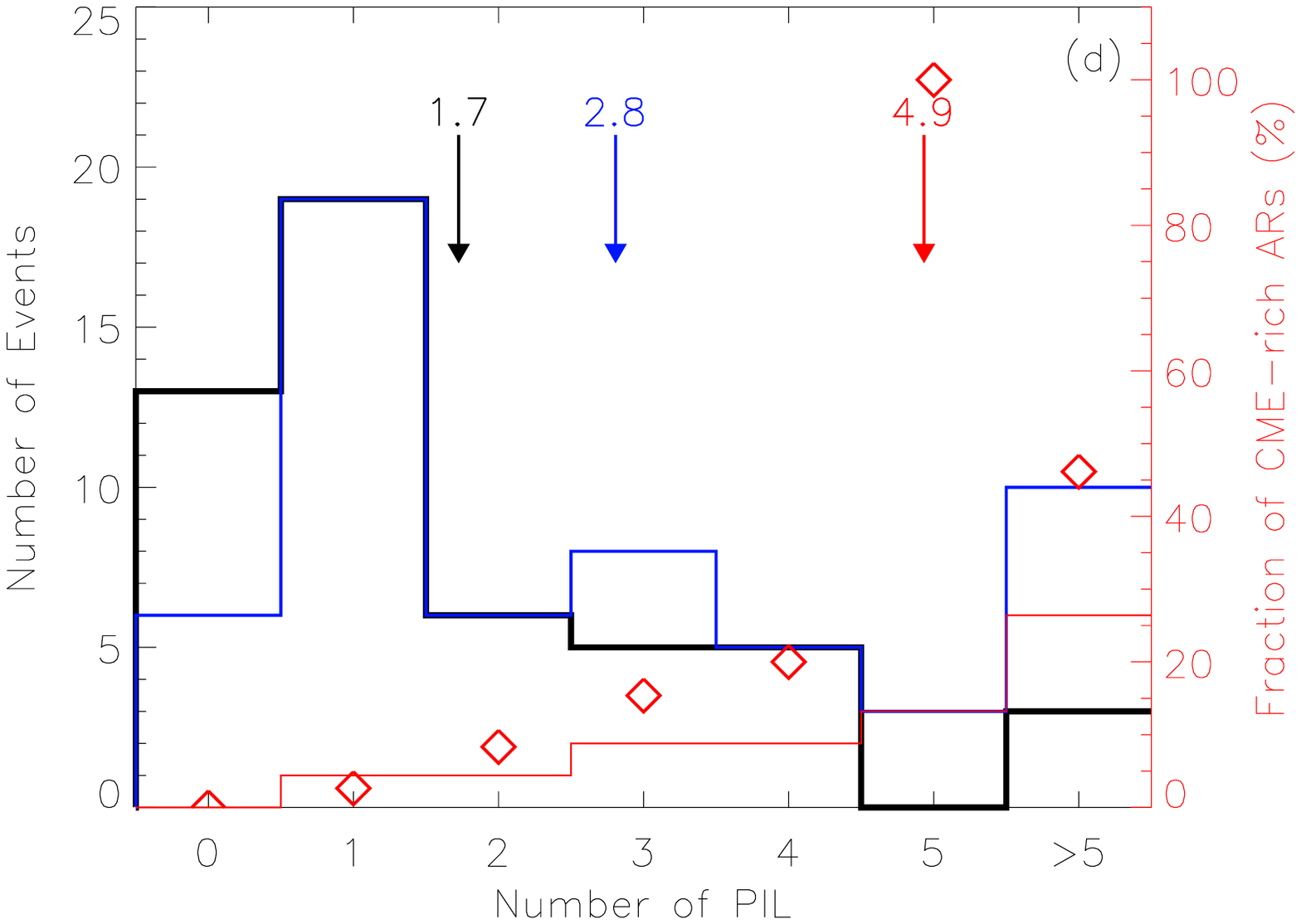} \\
  \caption{From panel (a) to (d), the histograms of the area ($A_t$), magnetic flux ($F_t$),
  length of PILs ($L_{pil}$), and number of PILs ($N_{pil}$) of CME-less (black),
  CME-producing (blue) and CME-rich (red) ARs are presented. The red
   diamond symbols denote the fraction of CME-rich ARs in all ARs.}\label{fg_ars}
\end{figure*}

Similar to what we did  before, we focus on the four AR parameters:
$A_t$, $F_t$, $L_{pil}$ and $N_{pil}$. Figure~\ref{fg_ars} presents
the distributions of the four AR parameters for the three different
types of ARs. The CME-less, CME-producing and CME-rich ARs are
plotted in black, blue and red color, respectively. It is clear that
the distributions  are different. A CME-producing AR tends to be
larger, stronger, and more complex than a CME-less AR. Generally,
all the average values of the four AR parameters for CME-producing
ARs are almost twice as large as those for CME-less ARs. Further,
CME-rich ARs have even larger values of the four parameters than
the other two types of ARs. The average values of $A_t$, $F_t$,
$L_{pil}$ and $N_{pil}$ for CME-rich ARs are about $12.91\times10^3$
Mm$^2$, $3.60\times10^{14}$ Wb, $120.3$ Mm and $4.9$, respectively,
which are 1.7, 1.7, 1.8 and 1.8 times  those of  CME-producing ARs,
and 2.4, 2.7, 3.6 and 2.9 times  those of  CME-less ARs. The
fraction  of the number of CME-rich ARs of  all ARs in each bin is
denoted by the red diamonds in Figure~\ref{fg_ars}. It can be found
that the  fraction of CME-rich ARs generally increases with the
increasing values of AR parameters. These results suggest that an AR
with a larger area, stronger magnetic field and more complex
morphology is more likely to be a CME-rich AR.

In particular, we notice that there is only one CME-rich AR with
$A_t\leq4000$ Mm$^2$, three CME-rich ARs with
$F_t\leq1.5\times10^{14}$ Wb, two CME-rich ARs with $L_{pil}\leq25$
Mm, one CME-rich AR with $N_{pil}\leq 1$, and further only one
CME-rich AR with all the above conditions satisfied. Thus, these
values, $A_t=4000$ Mm$^2$, $F_t=1.5\times10^{14}$ Wb, $L_{pil}=25$
Mm, and $N_{pil}=1$, can be treated as effective thresholds, below
which an AR is hard to frequently produce CMEs. Moreover, one PIL
implies that the AR has a dipole field, which is the most simple
topology of ARs on the Sun. Such ARs are not favorable for producing
multiple CMEs. In Figure 5 of our previous paper
\citep{Wang_Zhang_2008}, we  showed the distributions of the four
parameters  for all the 1143 MDI ARs during Carrington rotation 1911
-- 2051. By comparing these thresholds to the distributions, we find
that the value of each threshold is near the middle of its corresponding
distribution, which means that at each side of the thresholds there
are many ARs. Thus the values of these thresholds are meaningful in
distinguishing CME-rich ARs from others.

Further, we use a method called linear discriminant analysis (LDA)
to characterize two different classes of ARs, which have different CME productivity, in terms of these four
parameters. LDA is a widely used classification method in many
areas. Generally, LDA can be treated as a kind of special regression
analysis. One can refer to the paper by \citet{Fisher_1936} for its
principle, and refer to Sec.2.3 of our previous paper about solar
prominence recognition \citep{Wang_etal_2010} for more details of
its application. In this case, we have got four parameters for all
the 108 MDI ARs, and we also have known the CME productivity of these
ARs. Thus we can treat these ARs as a true table, and apply the 
LDA to derive the optimized combination of the four parameters for 
discriminating between any desired two classes of ARs with different CME 
productivity. The optimized combination 
of the parameters is called linear discriminant function (LDF) and
has the following form
\begin{eqnarray}
f=c_1\frac{A_t}{<A_t>}+c_2\frac{F_t}{<F_t>}+c_3\frac{L_{pil}}{<L_{pil}>}+c_4\frac{N_{pil}}{<N_{pil}>} \label{eq_2}
\end{eqnarray}
where $<x>$ indicates the average value of the quantity $x$ for 
all the 108 ARs used in our LDA, and $c_{1-4}$ are the coefficients.
The vector ($c_1$, $c_2$, $c_3$, $c_4$) defines a hyperplane in 
the four dimension space of the parameters ($A_{t}$, $F_{t}$, $L_{pil}$,
$N_{pil}$), which best separates the two classes of ARs. 
In a simplified form, the
optimum vector is achieved by the vector going from the mean values of the first class
to the mean value of the second one, while the practical computations also
involves the covariance of the distributions \citep{Fisher_1936}.

\begin{table*}[tb]
\begin{center}
\caption{Results of the linear discriminant analysis} \label{tb_lda}
\tabcolsep 25pt
\begin{tabular}{ccccccc}
\hline
&$c_1$ &$c_2$ &$c_3$ &$c_4$ &G \\
\hline
CME-less vs. -producing &-0.15 &0.20 &-0.26 &-0.07 &0.24 \\
CME-poor vs. -rich &-0.99 &0.87 &-0.26 &0.64 &0.76 \\
\hline
\end{tabular}
\end{center}
\footnotesize
$^*$ Column $c_{1-4}$ are the coefficients in Eq.\ref{eq_2}. The 
last column gives the goodness of LDF (see main text for details). 
The second and third row is for the discrimination between CME-less and 
CME-producing and between CME-poor and CME-rich, respectively.
\end{table*}

\begin{figure}[tb]
  \centering
  \includegraphics[width=\hsize]{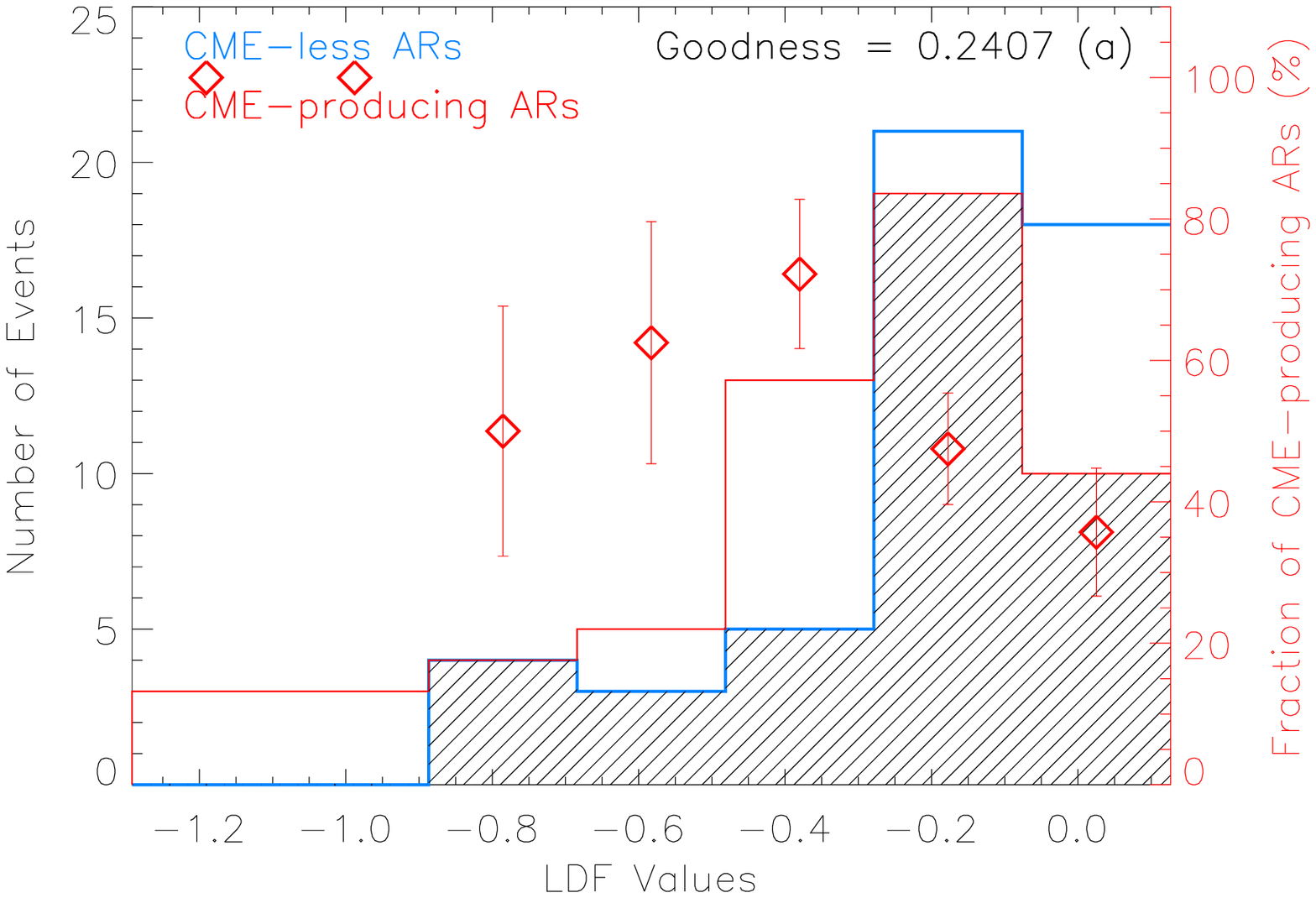}
  \includegraphics[width=\hsize]{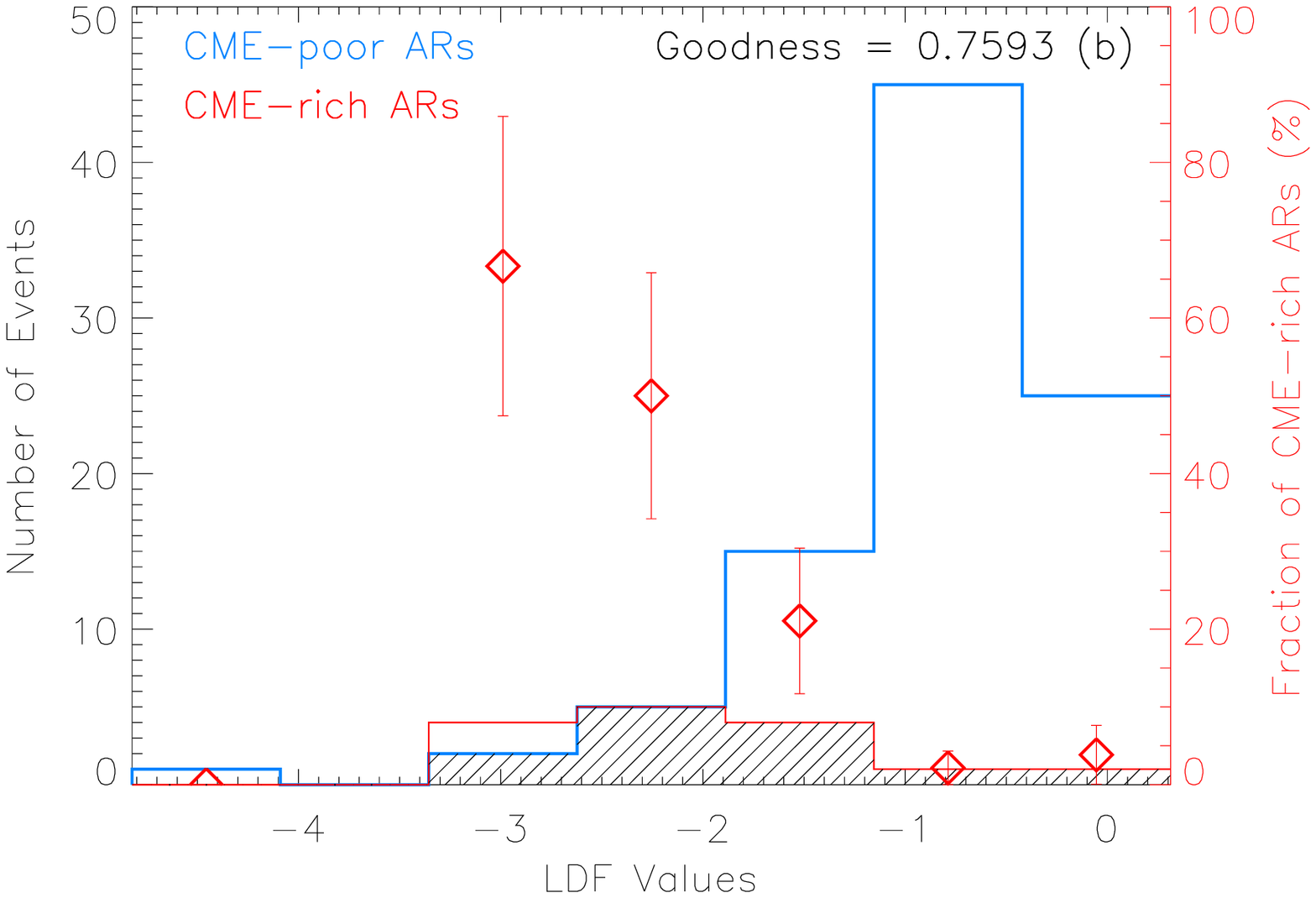}
  \caption{Histograms of LDF values for discriminating CME-less and CME-producing ARs (Panel a)
  and for discriminating CME-rich and CME-poor ARs (Panel b). The shadows represent the common
part of the two distributions (see the end of Sec.\ref{sec_productivity}).}\label{fg_lda}
\end{figure}

According to the LDF, one can get a one-dimensional distribution of the function value
$f$ for the two different classes of ARs (as seen in Fig.\ref{fg_lda}). 
As long as the distributions of the two different classes of ARs occupy 
different ranges of the $f$ value, the two classes of ARs can be more or less discriminated.
Here we try to derive two LDFs for discrimination between CME-less 
and CME-producing ARs, and between CME-poor (CME number less than 3) 
and CME-rich ARs, respectively. The derived optimized coefficients $c_{1-4}$
have been listed in Table~\ref{tb_lda}.

Figure~\ref{fg_lda} presents the LDA results. For discrimination 
between CME-less and CME-producing ARs (Fig.\ref{fg_lda}a), the overall goodness is
0.24. It is calculated by the formula
\begin{eqnarray}
G=1-\frac{n_0}{n}
\end{eqnarray}
where $n_0$ is the number of ARs whose LDF value falls within the 
overlap (indicated as the shadowed region in Fig.\ref{fg_lda}a) and 
$n$ is the total number of the ARs. $G=1$ means the LDF being 
able to completely discriminate between the two different classes of
ARs.
The fractions of CME-producing ARs marked by the red diamonds in
Figure~\ref{fg_lda}(a) suggest that about 69\% of ARs with LDF value
$\leq-0.3$ are CME-producing compared with the 43\% of ARs with LDF
value $>-0.3$, and particularly, all of ARs with LDF value
$\leq-0.9$ are CME-producing. The goodness of the discrimination for
CME-poor and CME-rich ARs is much better, which is 0.76
(Fig.\ref{fg_lda}(b)). On the left-hand side of the LDF value of
$-1.9$, the fraction of CME-rich ARs is about 53\%, while on its
right-hand side, the fraction is only about 7\%. Particularly, almost
all the ARs with LDF value $>-1.0$ cannot be a CME-rich AR.

\section{CME-rich ARs}\label{sec_CME-rich}
\subsection{Pace of CME Occurrence}
\begin{figure}[tb]
  \centering
  \includegraphics[width=\hsize]{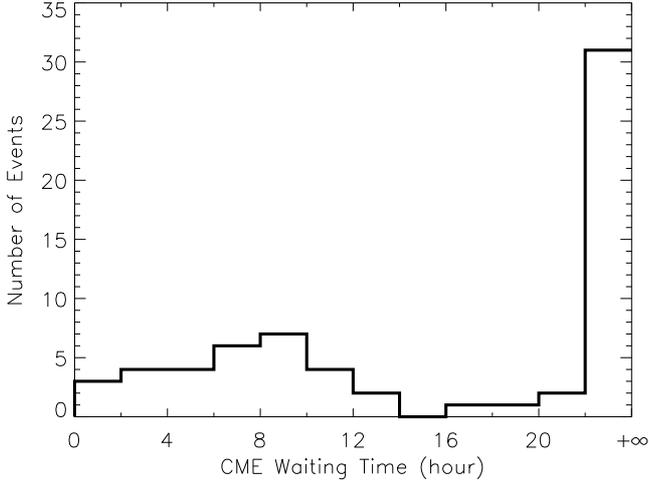}
  \caption{Distribution of the waiting times of same-AR CMEs. The first appearance of
CMEs in the field of view of LASCO C2 is adopted in calculating the waiting
time.}\label{fg_time}
\end{figure}

In certain  aspects, CME-rich ARs are  more interesting, especially
for the purpose of space weather prediction. In our sample, there
are a total of 15 CME-rich MDI ARs, which produced at least 80 CMEs.
During solar minima,  on average one CME occurs every other day
\citep[e.g.,][]{Gopalswamy_2006}. Thus, a question that naturally
rises  is how frequently
 CMEs take  place in  these CME-rich ARs. Here we call the CMEs from the same AR
same-AR CMEs. Figure~\ref{fg_time} shows the distribution of the
time interval ( so called waiting time) between two successive
same-AR CMEs for these 80 CMEs. The time of the first appearance
of CMEs in LASCO field of view is used to calculate the interval. It
is found that the distribution can be roughly divided into two
parts. The first part contains waiting times less than 15 hours and
the second part longer than 15 hours. For the second part, we simply
think that there is no tightly physical connection between two
successive same-AR CMEs, because of the longer time
interval. More attention will be put on the events of the first
part.  This part includes 30 data points, and manifests a unimodal
distribution with a peak around $8$ hours. It can be read from the
figure that about 43\% of the waiting times fall into the interval
of  $6-10$ hours, and about 83\% of them are between $2$ and $12$
hours. It is suggested that these successive same-AR CMEs usually
occur in a pace of about $8$ hours. We would like to call these CMEs
related same-AR CMEs. Few of such CMEs can take place within 2
hours or after 12 hours of a preceding CME.
A further discussion of the waiting time of the
related same-AR CMEs will be pursued in
Sec.\ref{sec_discussion}.

\begin{figure}[tb]
  \centering
  \includegraphics[width=\hsize]{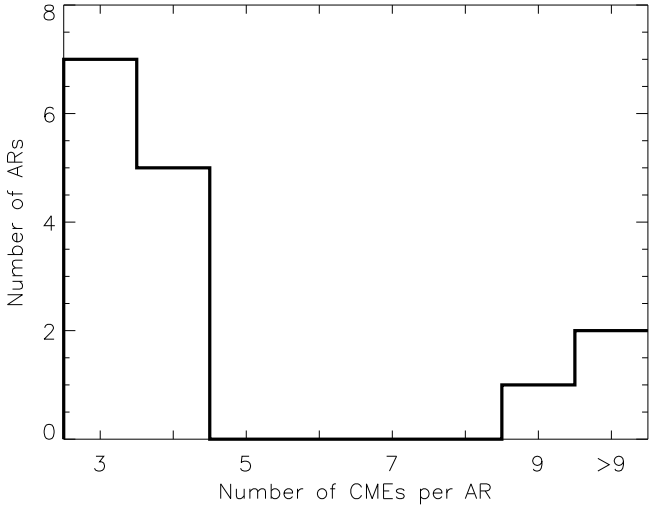}
  \caption{Histogram distribution of the number of CMEs produced by the CME-rich ARs.}\label{fg_cmeproductivity}
\end{figure}

Further, Figure~\ref{fg_cmeproductivity} shows the CME productivity
of these CME-rich ARs. It is found that there are actually three ARs
producing 9 or more CMEs, and all the rest had  produced 3 or 4
CMEs. The most productive AR had 19 CMEs (labeled as AR-a
hereafter),  which is NOAA AR 8210 appearing during Carrington
rotation 1935. The other two most productive ARs had 11 and 9 CMEs
(labeled as AR-b and AR-c), respectively. AR-b is NOAA AR 8100
appearing during Carrington rotation 1929 , while AR-c is a complex of 
NOAA AR 9395, 8398 and 8399 appearing during Carrington rotation 1943.
Table~\ref{tb_cme-rich} lists the three most productive ARs and
related CMEs.

\begin{table*}[p]
\begin{center}
\caption{Most productive ARs and corresponding CMEs} \label{tb_cme-rich}
\tabcolsep 10.3pt
\begin{tabular}{rrrrrrrl}
\hline
\bf For ARs & {\bf CR} & {\bf Location} & $\mathbf A_t$ & $\mathbf F_t$ & $\mathbf L_{pil}$ & $\mathbf N_{pil}$ & {\bf NOAA}  \\
 &  & deg & $\times10^3$ Mm$^2$ & $\times10^{14}$ Wb & Mm &  & AR \\
For CMEs & \multicolumn{2}{r}{Date Time} & Location & CPA & Width & Speed & \\
    & \multicolumn{2}{r}{UT}   &     & deg & deg & km s$^{-1}$ & \\
\hline
\bf AR-a & {\bf 1935} & {\bf (138, -17)} & {\bf 9.68} & {\bf 3.21} & {\bf 140} & {\bf 3} & {\bf 8210} (Middle) \\
     a1 & \multicolumn{2}{r}{1998/04/25 15:11} & S21E76 &  95 &  73 &  349 & \\
     a2 & \multicolumn{2}{r}{1998/04/25 18:38} & S13E73 &  70 &  17 &  324 & \\
     a3 & \multicolumn{2}{r}{1998/04/27 08:56} & S16E51 &  halo & 360 & 1385 & \\
     a4 & \multicolumn{2}{r}{1998/04/29 05:31} & S16E30 & 148 &  85 &  327 & \\
     a5 & \multicolumn{2}{r}{1998/04/29 16:58} & S15E19 &  halo & 360 & 1374 & \\
     a6 & \multicolumn{2}{r}{1998/05/01 23:40} & S19W02 &  halo & 360 &  585 & \\
     a7 & \multicolumn{2}{r}{1998/05/02 05:31} & S17W10 &  halo & 360 &  542 & \\
     a8 & \multicolumn{2}{r}{1998/05/02 14:06} & S14W15 &  halo & 360 &  938 & \\
     a9 & \multicolumn{2}{r}{1998/05/02 21:20} & S20W18 & 226 &  49 &  338 & \\
    a10 & \multicolumn{2}{r}{1998/05/03 10:29} & S14W31 & 241 &  74 &  497 & \\
    a11 & \multicolumn{2}{r}{1998/05/03 22:02} & S15W35 & 317 & 194 &  649 & \\
    a12 & \multicolumn{2}{r}{1998/05/04 00:58} & S14W41 & 270 &  66 &  279 & \\
    a13 & \multicolumn{2}{r}{1998/05/04 23:27} & S20W43 & 240 &  39 &  338 & \\
    a14 & \multicolumn{2}{r}{1998/05/05 00:58} & S13W48 & 319 &  60 &  218 & \\
    a15 & \multicolumn{2}{r}{1998/05/06 00:02} & S21W59 & 274 & 110 &  786 & \\
    a16 & \multicolumn{2}{r}{1998/05/06 08:29} & S15W67 & 309 & 190 & 1099 & \\
    a17 & \multicolumn{2}{r}{1998/05/06 09:32} & S13W75 & 264 &  95 &  792 & \\
    a18 & \multicolumn{2}{r}{1998/05/07 11:05} & S15W80 & 270 &  16 &  483 & \\
    a19 & \multicolumn{2}{r}{1998/05/08 14:32} & S16W89 & 259 &  80 &  777 & \\
\bf AR-b & \bf 1929 & \bf (351, -20) & \bf 8.31 & \bf 2.64 & \bf 147 & \bf 7 & {\bf 8100} (Emerging)\\
     b1 & \multicolumn{2}{r}{1997/10/29 18:21} & S19E45 &  88 &  62 &  133 & \\
     b2 & \multicolumn{2}{r}{1997/11/03 05:28} & S16W20 & 240 & 109 &  227 & \\
     b3 & \multicolumn{2}{r}{1997/11/03 09:53} & S14W18 & 238 &  71 &  338 & \\
     b4 & \multicolumn{2}{r}{1997/11/03 11:11} & S13W23 & 233 & 122 &  352 & \\
     b5 & \multicolumn{2}{r}{1997/11/04 06:10} & S15W32 &  halo & 360 &  785 & \\
     b6 & \multicolumn{2}{r}{1997/11/04 15:50} & S18W32 & 242 &   5 &  266 & \\
     b7 & \multicolumn{2}{r}{1997/11/05 04:20} & S15W46 & 264 &  49 &  271 & \\
     b8 & \multicolumn{2}{r}{1997/11/05 07:29} & S16W49 & 287 &  40 &  350 & \\
     b9 & \multicolumn{2}{r}{1997/11/05 12:10} & S15W50 & 270 &  52 &  356 & \\
    b10 & \multicolumn{2}{r}{1997/11/06 12:10} & S17W62 &  halo & 360 & 1556 & \\
    b11 & \multicolumn{2}{r}{1997/11/08 08:59} & S17W88 & 271 &  76 &  453 & \\
\bf AR-c & \bf 1943 & \bf (182, 20)  & \bf 35.61 & \bf 9.14 & \bf 122 & \bf 5 & \bf 8395, 8398, 8399 \\
 &  &  &  &  &  &  & (Decaying) \\
     c1 & \multicolumn{2}{r}{1998/11/24 13:23} & N26E84 &  54 &  50 &  248 & \\
     c2 & \multicolumn{2}{r}{1998/11/24 23:30} & N32E78 &  50 &  61 &  432 & \\
     c3 & \multicolumn{2}{r}{1998/11/25 06:30} & N18E72 &  53 &  41 &  256 & \\
     c4 & \multicolumn{2}{r}{1998/11/25 14:30} & N20E73 &  57 &  52 &  213 & \\
     c5 & \multicolumn{2}{r}{1998/11/26 11:30} & N19E57 &  45 &  50 &  216 & \\
     c6 & \multicolumn{2}{r}{1998/11/28 06:30} & N20E46 &  62 &  88 &  495 & \\
     c7 & \multicolumn{2}{r}{1998/12/05 19:32} & N33W40 & 340 &  23 &  --- & \\
     c8 & \multicolumn{2}{r}{1998/12/06 03:54} & N34W46 & 331 &  36 &  159 & \\
     c9 & \multicolumn{2}{r}{1998/12/07 15:30} & N28W62 & 327 &  42 &  490 & \\
\hline
\end{tabular}
\end{center}
\footnotesize
$^*$ The table lists the information of each most productive AR (bold fonts) with the 
corresponding CMEs (normal fonts) in the following rows. The first column numbers the 
ARs and CMEs. For ARs, the other columns from the left to right are Carrington Rotation (CR),
Location in Carrington coordinates, area ($A_t$), magnetic flux ($F_t$), length and
number of PILs ($L_{pil}$ and $N_{pil}$), and the corresponding NOAA AR with its phase
indicated in parentheses. For CMEs,
the other columns give the date, time, location in heliographic coordinates, central
position angle ($CPA$), apparent width and speed.
\end{table*}

\begin{figure}[tbh]
  \centering
  \includegraphics[width=0.72\hsize]{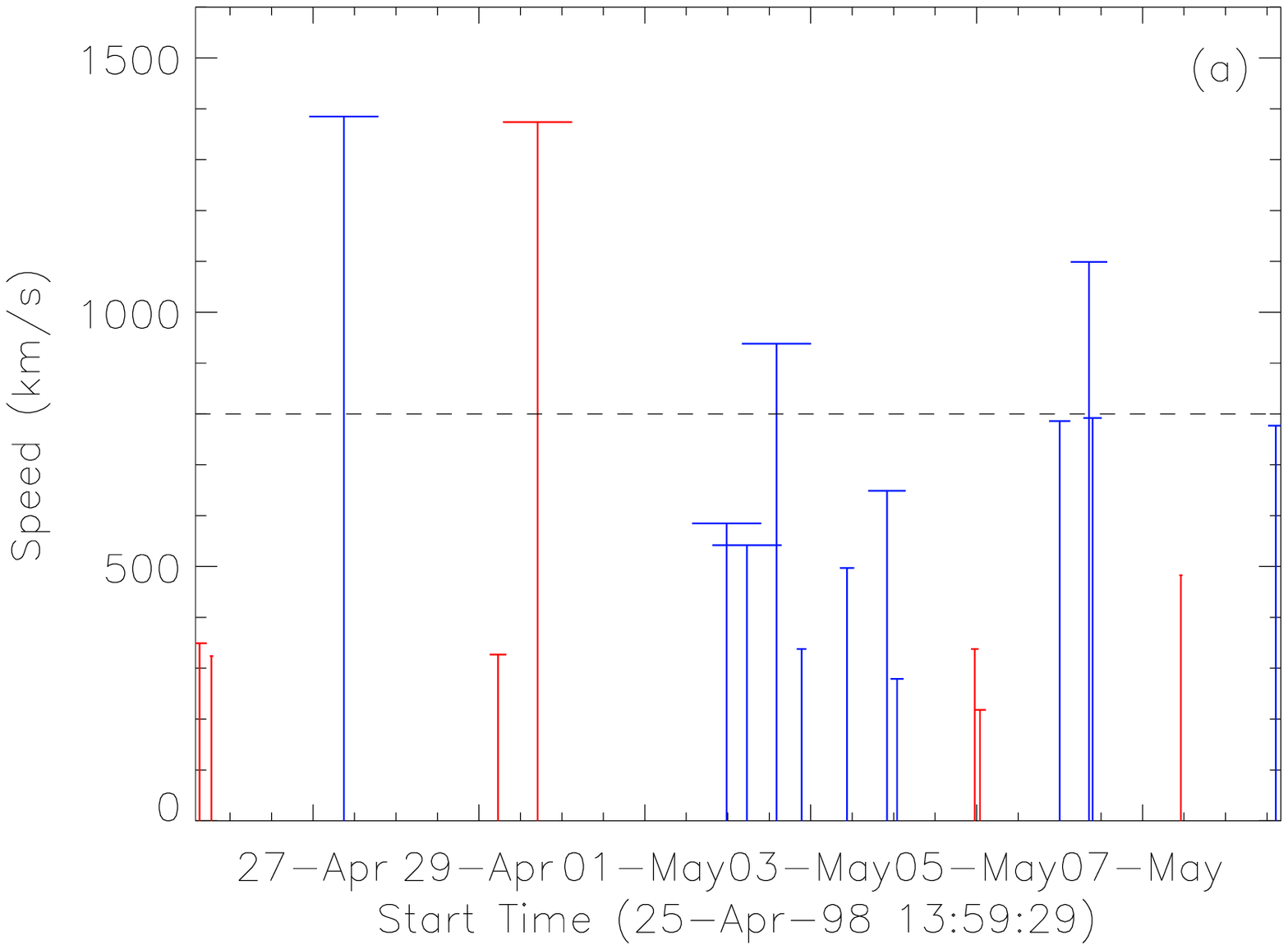}
  \includegraphics[width=0.72\hsize]{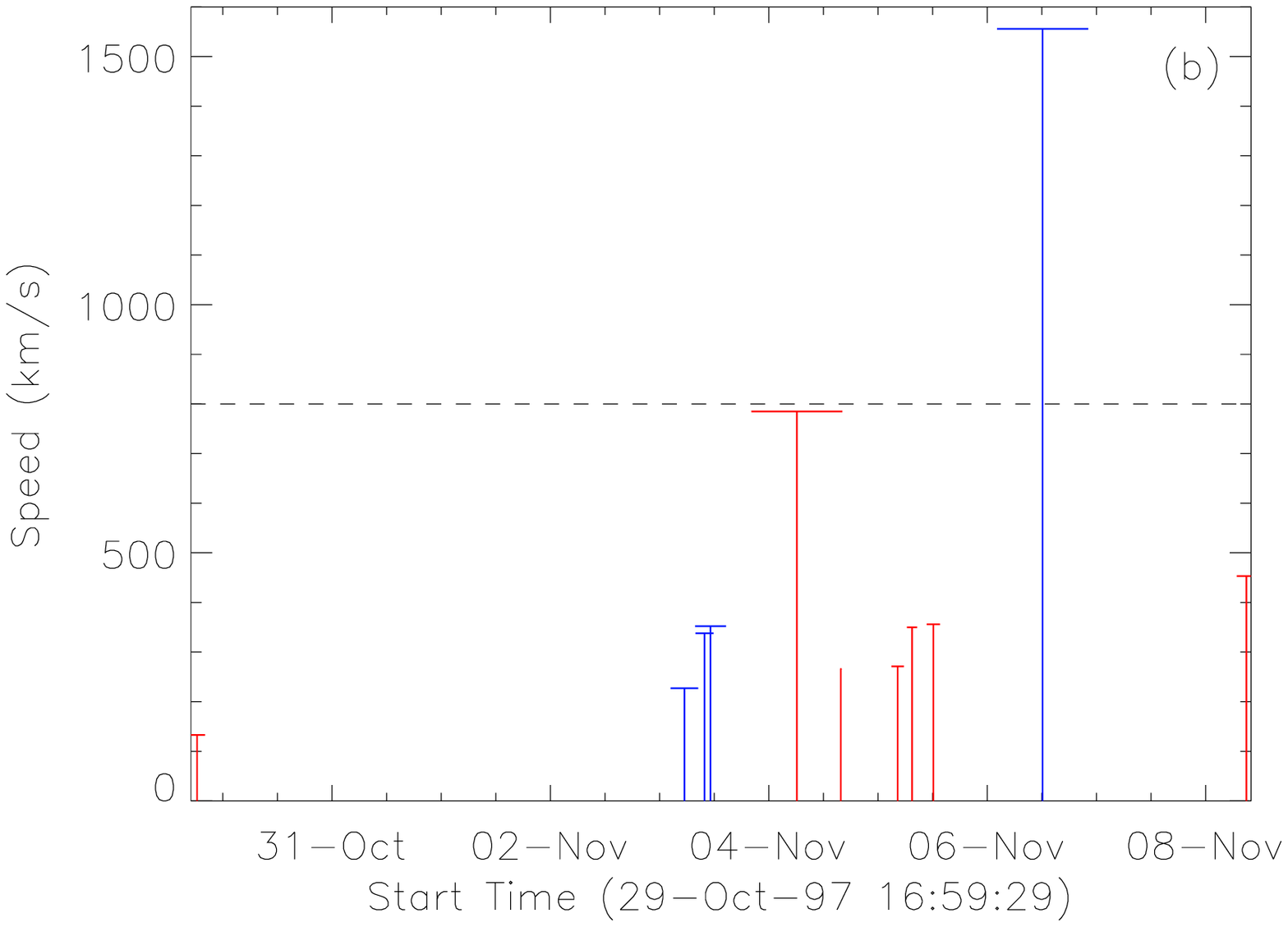}
  \includegraphics[width=0.72\hsize]{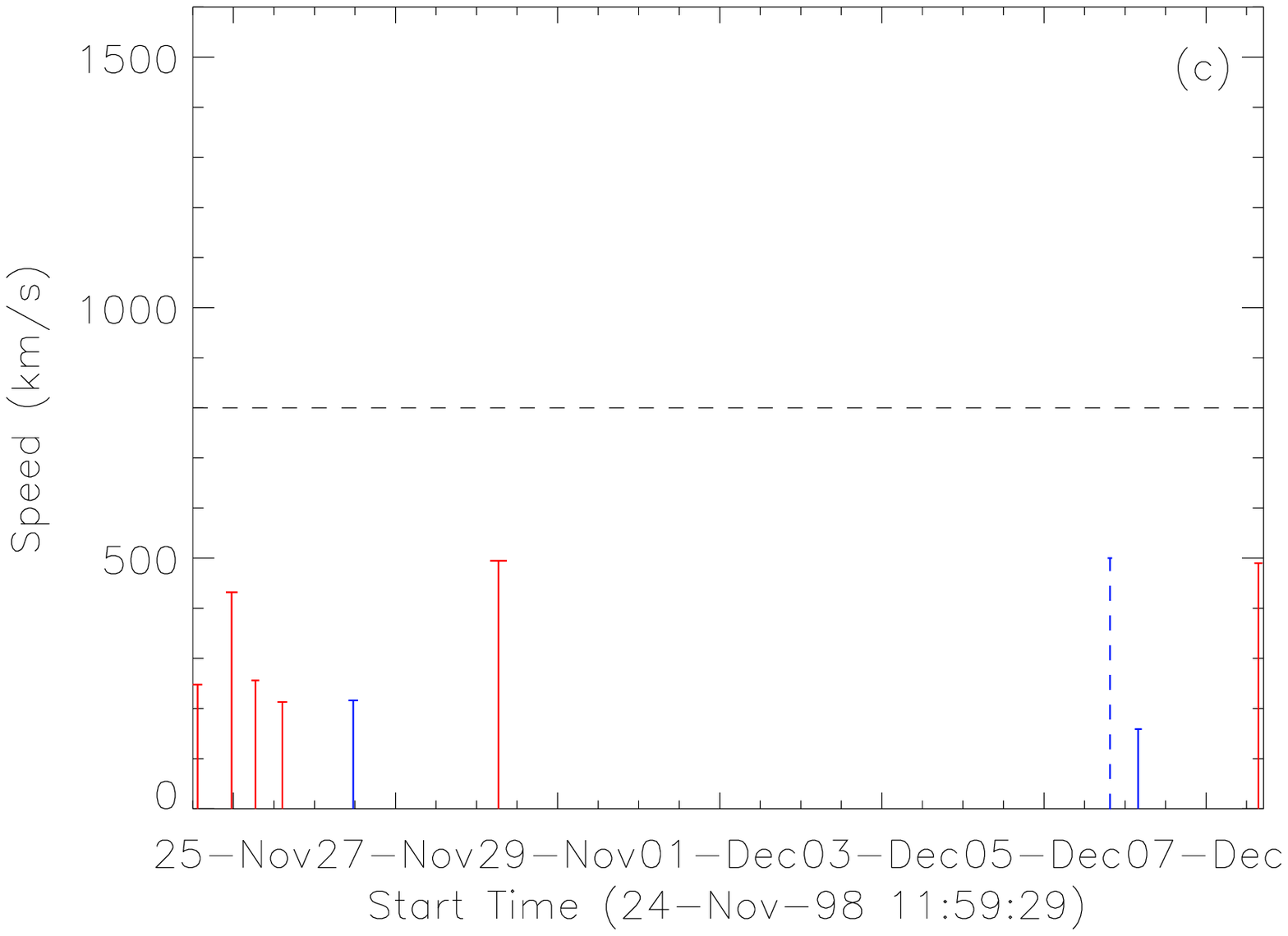}
  \caption{Panel (a) -- (c) present the associated CMEs of three most productive ARs: AR-a, AR-b and AR-c, respectively.
Each vertical line stands for a CME, and its length indicates the CME
apparent  speed. The horizontal bar at the top of each line indicates the CME
angular width. The longer the bar is, the wider is the CME's angular span. 
Alternating color is used to group the related same-AR CMEs,
among which the waiting times between CMEs are no more than 15
hours. Horizontal dashed line marks the speed of 800 km s$^{-1}$. In
Panel (c), the dashed vertical line indicates a CME without an
effective  speed.}\label{fg_productive_ars}
\end{figure}

The frequency of CME occurrence of these ARs is illustrated in
Figure~\ref{fg_productive_ars}. Each vertical line in the plots
stands for a CME. Its length indicates the CME apparent  speed and
the horizontal bar at the top indicates the width. The lines with
the same color mean that these CMEs are related, i.e.,
the time interval between two successive CMEs is shorter than 15
hours. For AR-a there are 8 groups (indicated by
alternating colors of red and blue) of related same-AR CMEs, and
for AR-b and AR-c there are 5 groups each. A first impression
obtained from these plots is that there is only one CME that can be
faster than 800 km/s in any one group, and 2 out of 3 extremely fast
CMEs ($>1200$ km s$^{-1}$) were isolated (the other one was only
grouped with another slow CME). The other 12 CME-rich ARs all follow
the above regulation (not shown in the figure). Since CME speed can
be used as a proxy of CME energy, or the free energy released from
ARs, we simply treat a CME faster than 800 km s$^{-1}$ as a strong
CME, and others as weak CMEs. The above facts imply that (1) the
total free magnetic energy stored in an AR at any instant can
usually  support at most one strong CME and several weak CMEs, and
(2)  an AR has to take more than 15 hours to re-accumulate
sufficient free energy to produce another strong CME.

\begin{table*}[tbh]
\begin{center}
\caption{Selected CME-less ARs} \label{tb_cme-less}
\tabcolsep 13.1pt
\begin{tabular}{rrrrrrrl}
\hline
No & {CR} & {Location} & $A_t$ & $F_t$ & $L_{pil}$ & $N_{pil}$ & {NOAA}  \\
 &  & deg & $\times10^3$ Mm$^2$ & $\times10^{14}$ Wb & Mm &  & AR \\
\hline
 1 & 1920 & (205,   7) &  8.17 &  1.67 &   0 &  0 & 8020 ($\beta$)\\
 2 & 1922 & ( 14,   5) &  4.21 &  1.43 &  62 &  2 & 8040 ($\beta$)\\
 3 & 1923 & (188, -28) &  3.86 &  1.01 &  17 &  1 & 8048 ($\beta$)\\
 4 & 1926 & (268,  26) &  6.17 &  1.18 &   0 &  0 & 8074 ($\alpha$)\\
 5 & 1926 & (279,  16) &  2.48 &  0.54 &   0 &  0 & 8073 ($\alpha$)\\
 6 & 1926 & ( 11,  34) &  3.10 &  0.61 &   6 &  1 & 8081 ($\alpha$)\\
 7 & 1927 & (225,  28) &  8.19 &  2.01 &  15 &  1 & 8086 ($\beta$)\\
 8 & 1927 & ( 97, -24) &  6.84 &  1.19 &   8 &  1 & 8087 ($\alpha$)\\
 9 & 1927 & (363,  22) &  4.48 &  1.08 &  18 &  2 & 8082 ($\beta$)\\
10 & 1928 & (342, -30) &  2.79 &  0.52 &  18 &  1 & 8090 ($\alpha$)\\
11 & 1928 & ( 22,  18) &  2.91 &  0.61 &  25 &  3 & 8099 ($\beta$)\\
12 & 1929 & (303,  23) &  4.88 &  1.08 &  60 &  3 & 8103 ($\beta$)\\
13 & 1929 & ( 91, -19) &  2.67 &  0.60 &  11 &  1 & 8109 ($\beta$)\\
14 & 1930 & (352, -20) & 14.01 &  2.68 &   0 &  0 & 8112 ($\alpha$)\\
15 & 1930 & (358,  25) &  2.79 &  0.52 &   0 &  0 & 8111 ($\alpha$)\\
16 & 1930 & (287, -29) &  1.26 &  0.25 &   0 &  0 & 8114 ($\beta$)\\
17 & 1931 & (345, -23) & 13.68 &  3.75 & 156 &  2 & 8124 ($\beta\gamma$)\\
18 & 1932 & (278, -37) &  6.66 &  1.82 &  72 &  4 & 8143 ($\beta\gamma$)\\
19 & 1932 & ( 14, -20) &  3.31 &  0.63 &  10 &  1 & 8158 ($\beta$)\\
20 & 1932 & (267,  14) &  3.55 &  0.89 &   0 &  0 & 8144 ($\beta$)\\
21 & 1932 & ( 24,  26) &  2.47 &  0.57 &   0 &  0 & 8157 ($\alpha$)\\
22 & 1933 & ( 62, -40) &  8.16 &  2.26 &  69 &  4 & 8176 ($\beta$)\\
23 & 1933 & (360,  22) &  3.97 &  0.72 &  16 &  1 & 8160 ($\beta$)\\
24 & 1934 & (240, -24) & 18.22 &  4.93 & 161 &  6 & 8185, 8189 ($\beta\gamma$)\\
25 & 1934 & ( 83, -23) &  8.25 &  2.76 &  48 &  3 & 8193, 8199 ($\beta$)\\
26 & 1935 & (386, -23) & 40.52 & 10.17 & 151 & 10 & 8195, 8194, 8198, 8200, \\
 &  &  &  &  &  &  & 8202 ($\beta$) \\
27 & 1935 & (356,  18) &  2.65 &  0.48 &  11 &  1 & 8201 ($\alpha$)\\
28 & 1936 & (282,  22) & 12.97 &  3.31 &  92 &  4 & 8222 ($\beta$)\\
29 & 1936 & (282, -27) &  8.65 &  2.27 &  85 &  4 & 8220 ($\beta$)\\
30 & 1937 & (283,  22) &  5.30 &  0.92 &   0 &  0 & 8238, 8239 ($\beta$)\\
\hline
\end{tabular}
\end{center}
\footnotesize
$^*$ The column arrangement is as the same as that for ARs in Table \ref{tb_cme-rich}
except that the parentheses in the last column give the most complicated 
type of the AR associated sunspot group during the AR crossing the visible 
disk.
\end{table*}

\citet{Kienreich_etal_2011} reported four homologous CME-associated
coronal waves observed by STEREO. It is found that the waiting times
between the eruptions have a positive correlation with the strength
of the eruptions. This case study suggests that from the same AR a
stronger eruption needs a longer waiting time, which is consistent
with our statistical results.

\subsection{Most Productive ARs vs. CME-less ARs}\label{sec_rich-less}

 MDI daily magnetogram images  indicate that all the three CME-productive ARs discussed above rotated
from the solar east limb to west limb, and lasted at least for about
13 days. AR-a, i.e., NOAA AR 8210, has been studied by several
researchers. \citet{Subramanian_Dere_2001} pointed out that the life
time of this AR is about $65-79$ days, and it was in the mid-phase
when it appeared in the front-side of the Sun during Carrington
rotation 1935.
 The type of the sunspots associated with this AR changed among
$\beta\gamma$, $\beta\delta$, $\gamma\delta$ and
$\beta\gamma\delta$, indicating its complexity in morphology. AR-b
is also a complex AR. Different from AR-a, it was obviously emerging
on its way crossing the field of view. Its associated sunspots
developed from type of $\beta$ to $\beta\gamma$ and
$\beta\gamma\delta$ around 1997 November 2 -- 4, during and after which all the CMEs 
except one launched. AR-c
was  more complicated than AR-a and AR-b, which consisted of three
NOAA ARs. Our AR-detection method merges the three NOAA ARs together
as a single compound region, as it is indeed difficult to separate
them as viewed in magnetograms (an AR appears much bigger in the
magnetogram images than in the white light image). AR-c was probably
in the decaying phase. From MDI magnetograms, one may notice that
this AR was much more diffusive than other two. The average magnetic
field of AR-a and AR-b was larger than 300 G, whereas that of AR-c
was about 250 G. There were several sunspot groups in the AR, but
their types are $\beta$ or $\beta\gamma$, relatively simpler than
those in other two ARs. Thus, AR-c was a globally complex, but
locally simple and weak AR. This is probably why AR-c produced 9 CMEs but
none of these CMEs was faster than 500 km s$^{-1}$.

As a comparison, we look into   CME-less ARs. It is found that 19
out of 51 ($\sim37\%$) CME-less ARs have more than one PILs, and
only 4 ($\sim8\%$) CME-less ARs have the PILs' total length longer
than 100 Mm. Further, we checked the MDI magnetograms and NOAA AR
list, and selected  the CME-less ARs that have corresponding NOAA
ARs and showed in rotation  from the solar east limb to west limb.
There are a total of 30 such CME-less ARs. Table~\ref{tb_cme-less}
lists these ARs for reference. The sunspot classification suggests
that about 90\% of these ARs are very simple, belonging to $\alpha$
or $\beta$ type, and the other 10\% are $\beta\gamma$. Note that the
sunspot type we provide here is the most complex type during its
passage. We also investigated the MDI movies, and found that most of
these ARs are in the mid-phase of its whole life, and some in
emerging phase and others  in decaying phase. Compared with the most
productive ARs, the above results suggest that the CME productivity
of ARs is strongly related with the AR complexity, but less related
with the AR phase.

\section{Conclusions}\label{sec_conclusions}
In this paper, 224 location-identified CMEs and the corresponding
108 MDI ARs during 1997 -- 1998 are investigated. The association
between  CMEs and ARs suggests that about $63\%$ of the CMEs are related
with ARs, and at least about $53\%$ of the ARs
 produce one or more CME during one disk passage. Some ARs frequently produce CMEs; there are  15
CME-rich ARs, which produced a total of at least 80 CMEs, and the
most productive AR produced 19 CMEs. By analyzing the relationship
between the properties of CMEs and ARs, the following conclusions
are reached. These conclusions mostly confirm the previous studies
\citep[e.g.,][]{Guo_etal_2007, Falconer_etal_2008, Yeates_etal_2010} but
with significant additions.

\begin{enumerate}
\item There is no evident difference between AR-related and non-AR-related
CMEs in terms of CME speed, acceleration and width, which suggests
that the concept of two types of CMEs \citep[e.g.,][]{Sheeley_etal_1999} may not be true,
or at least they can not be simply attributed to their source regions.

\item There is no evident dependence of CME speed on the AR area, magnetic
flux and complexity, though a trend that an AR with larger area,
stronger magnetic field and more complex morphology has a higher
possibility of producing extremely fast CMEs (speed $>1500$ km
s$^{-1}$) was found before \citep{Wang_Zhang_2008}. However, the CME
width manifests a weak correlation with the AR parameters, and the
area and magnetic flux are two  important factors.

\item CME-producing ARs more likely appear in the two latitudinal belts at
$\pm(15^\circ-30^\circ)$ than CME-less ARs. Particularly, all
CME-rich ARs are located in the belts, and only 18\% of the
ARs outside the two belts can produce CMEs.

\item CME-producing ARs tend to be larger, stronger and more complex than
CME-less ARs. All the average values of $A_t$, $F_t$, $L_{pil}$ and
$N_{pil}$ of CME-producing ARs are almost twice as large as those
of CME-less ARs. For CME-rich ARs, the average values are even
larger, which are 2.4, 2.7, 3.6 and 2.9 times  those of  CME-less
ARs.

\item There seem to be thresholds of $A_t=4000$ Mm$^2$, $F_t=1.5\times10^{14}$ Wb
and $L_{pil}=25$ Mm, below which an AR is hard to
frequently produce CMEs. Particularly, a dipolar-field AR is not favorable
for producing multiple CMEs. The discriminant analysis shows that almost
all the ARs with the LDF value larger than $-1.0$ cannot be a
CME-rich AR.

\item The sunspots in all the three most productive ARs (creating 9 or more CMEs) at
least belong to $\beta\gamma$ type, whereas 90\% of those in the
CME-less ARs are $\alpha$ or $\beta$ type, and only 10\%
$\beta\gamma$ type. It is suggested that the CME productivity of ARs is strongly related
with the AR complexity, but less related with its phase.

\item Combining the above results, we can claim that the size, strength
and complexity of ARs do little with the kinematic properties of CMEs,
but have significant effects on the CME productivity.
\end{enumerate}

The CME-rich ARs are then investigated particularly. Through the analysis
of the waiting times of the same-AR CMEs, it is found that the
distribution of the waiting times consists of two parts with a
separation at about 15 hours, which implies two different patterns
of the occurrences of same-AR CMEs, and those CMEs with a waiting
time shorter than 15 hours are probably truly physical related. A
detailed analysis of these related same-AR CMEs further gives rise  to
the following two interesting conclusions.

\begin{enumerate}
\item The average waiting time of related same-AR CMEs is about 8 hours,
which means that a CME-productive AR tends to produce CMEs at a pace
of 8 hours.

\item An AR cannot produce two or more CMEs faster than 800 km s$^{-1}$
within a time interval of 15 hours (i.e., in any group of
related same-AR CMEs).
\end{enumerate}

It should be noted that all the above conclusions are established 
on the statistical study of CMEs and ARs near the minimum of solar 
cycle 23. Whether or not they also reflect the fact during solar 
maximum needs to be verified by further work.

\section{Preliminary Discussion On The CME Waiting Time}\label{sec_discussion}
A CME is a process of releasing  a huge amount of free magnetic
energy stored in the corona. Sufficient amount of free magnetic energy
is a necessary condition for an AR to produce a CME
\citep[e.g.,][]{Priest_Forbes_2002, Regnier_Priest_2007}. Many
previous studies also suggested that sufficient large helicity
injection is critical for a solar eruption
\citep[e.g.,][]{Demoulin_etal_2002, Nindos_Zhang_2002,
Nindos_etal_2003, Green_etal_2002, Green_etal_2003,
LaBonte_etal_2007, Smyrli_etal_2010}. Our statistical analysis
results of CME waiting times naturally raise two issues. One (labeled as I1) is why
CME-rich ARs frequently produce CMEs, especially why in a pace of
about 8 hours. The other (labeled as I2) is why there can be at most one strong CME
(speed $>800$ km s$^{-1}$) in any group of related same-AR CMEs
or within an interval of 15 hours? Note, the value of speed 800 km
s$^{-1}$ is underestimated because of the projection effect.
Moreover, we believe that the values of 8 hours, 15 hours and 800 km
s$^{-1}$, might slightly vary if more CME-rich ARs during solar
maximum are included in the statistical sample. No matter what the exact
values are, to satisfactorily address the two issues, we need much more
work. The unprecedented data from SDO mission, which have much
higher resolution in both space and time than SOHO data, may help us
deepening our understanding of the nature of same-AR CMEs. Here,
we would like to carry out a preliminary discussion on the two
issues. For issue I1, we think that it implies at least three
possible mechanisms of the related same-AR CMEs. 

(1) The related same-AR CMEs come from the same part of an AR. The AR is
able to quickly refill enough free energy or helicity after  it is
consumed by a CME, so that multiple CMEs can be launched from the
same place. In this scenario, our statistical results imply that the
time-scale of the refilling is about 8 hours.
\citet{LaBonte_etal_2007} surveyed 48 X-class flare-producing
regions and found that these regions consistently had a larger
helicity change than non-flaring regions. Particularly, they found
that most of the X-flare regions can accumulate helicity for a CME
in a few days to a few hours. For example, the typical time of
helicity injection for NOAA AR 10486 to repeatedly produce CMEs is
about 10 hours. \citet{Kienreich_etal_2011} reported four homologous
CME-associated coronal waves observed by STEREO. The waiting times
between them are around 2.5 hours, and it is found that the waiting
time has a positive correlation with the strength of the eruption.
However, more events show a much longer waiting time. Also in the
paper by \citet{LaBonte_etal_2007}, the waiting time for NOAA AR
10720 is about 19 hours. \citet{Li_etal_2010} study of the
homologous CMEs during 1997 May 5 --16 showed that sufficient
energy is built up on the order of several days. Homologous CMEs 
not only originate from the same source region but also have the 
similar morphology. They can be considered as a special type of 
same-AR CMEs. We suggest that such long-waiting-time CMEs in 
\citet{Li_etal_2010} study should belong to the second part of our
distribution (Fig.\ref{fg_time}), and probably have a different
cause.

(2) There are several magnetic flux systems in the AR, which are all
possible to develop into a CME, and the eruption of one of them may
cause others unstable and eventually erupting. In this scenario, the
time-scale of the unstabilization caused by the preceding CME is
typically 8 hours. The MHD numerical simulation by
\citet{Peng_Hu_2007} provided such possibility in theory. In their
simulation, multi-polar magnetic configuration, which contains three
arcade systems, is set, and shearing motions are introduced to build
up free energy. It is found that an arcade may form a flux rope and
then erupt by the shearing motion of its adjacent arcades. The study
of the two successive CMEs originating from NOAA AR 10808 on 2005
September 13 by \citet{Liu_etal_2009} is an observational evidence.
Their analysis suggested that the launch of the second CME was
contributed by the first CME which partially removed the overlying
magnetic fields in the northern part of the AR.

(3) The related same-AR CMEs might come from the different parts
of the same magnetic flux system in the AR. The eruption of one part
may cause the other parts further erupting. This scenario is similar to
but not same as the second one, and the time-scale of unstabilization is also
required to be about 8 hours. An observational case supporting it is
the 2005 May 13 CMEs studied by \citet{Dasso_etal_2009}. In their
work, they found that the giant ICME observed by ACE on May 15
actually consisted of two magnetic clouds, which were corresponding
to two CMEs originating from NOAA AR 10759 on May 13. The much more
detailed multi-wavelength analysis further showed that the two CMEs
were formed from the magnetic fields above the different portion of
the same filament (or PIL), and the waiting time is about 4 hours.
There are also some other studies showing that different portions of
the same filament may erupt successively
\citep[e.g.,][]{Maltagliati_etal_2006, Gibson_etal_2006,
Liu_etal_2008}.

Which one is most likely to work for the related same-AR CMEs? To
answer this question, we need to carefully check the erupting
process of each CME with multiple-wavelength data, especially the
exact locations that the CMEs originate. This will be done in a
separate  paper.

For issue I2, we think that the key point is the rate of
free energy accumulation. According to previous statistical studies
\citep[e.g.,][]{Vourlidas_etal_2000}, the mass of a CME  is
typically $10^{12}$ kg. Thus a speed of 800 km s$^{-1}$ corresponds
to a kinetic energy of $3\times10^{23}$ J. It is also showed that
the injected thermal energy during a CME is on the same order of its
kinetic energy \citep[e.g.,][]{Akmal_etal_2001,
Ciaravella_etal_2001, Rakowski_etal_2007}. In our study, CME speeds
were measured in the field of view of SOHO/LASCO, which is beyond
$2R_S$. Thus the gravitational potential energy of a CME is
considerable, which can be estimated as about $2\times10^{23}$ J
under the assumption of the CME mass equal to $10^{12}$ kg and moved
from the heliocentric distance $1R_S$ to beyond $5R_S$. The sum of thermal,
kinetic and potential energies meet the minimum requirement of the
free energy for an AR to produce a
 CME with a speed of 800 km s$^{-1}$. The actual free energy released during
the CME should also include radiation energy, like flares. Relating
the minimum required free energy with the waiting time of at least
15 hours, we can estimate that the rate of an AR accumulating free
energy is on the order of $10^{19}$ J s$^{-1}$. This value is a very
coarse estimation, because  CME mass, speed and waiting time are all
very different case by case.

Recently, \citet{Li_etal_2011} proposed a so-called `twin-CME' scenario
to explain ground level events (GLEs). In their model, they found that
two CMEs successively erupting from the same (or nearby) AR in 8.7 hours 
are favorable for the generation of GLEs. The duration of 8.7 hours
represents the characteristic time for a turbulence decayed away. Their 
scenario is apparently supported by the GLEs observations in solar cycle 
23 (Table 1 in their paper). Does the number 8.7 have any underlying 
physical relationship with our 8 hours? It is worthy of follow-up studies.

In short, we would like to highlight the values, 8 hours, 15 hours,
800 km s$^{-1}$ and $10^{19}$ J s$^{-1}$ derived/estimated
from our statistical study. These values can serve as constraints
for AR and/or CME modeling, and further deepen our understanding of
the mechanism of AR energy accumulation and release.

\acknowledgments{
We acknowledge the use of the data from SOHO/MDI and the CDAW CME
catalog, which is generated and maintained at the CDAW Data Center
by NASA and The Catholic University of America in cooperation with
the Naval Research Laboratory. SOHO is a project of international
cooperation between ESA and NASA. We thank the anonymous referees
for their kindly comments and corrections. This research is supported by
grants from 973 key project 2011CB811403, NSFC 41131065, 40904046,
40874075, 41121003, CAS 100-Talent Program,
KZCX2-YW-QN511 and startup fund, FANEDD 200530, and the
fundamental research funds for the central universities. J.Z. was 
supported by NSD grant ATM-0748003 and NASA grant NNG05GG19G.
}

\bibliographystyle{agufull}
\bibliography{../../../ahareference}

\end{document}